\documentclass[a4paper,11pt]{article}
\pdfoutput=1 

\usepackage{jinstpub} 
\usepackage{lineno,hyperref}
\usepackage{comment}
\usepackage{color}
\usepackage{todonotes}
\usepackage{siunitx}
\usepackage{booktabs}
\usepackage{adjustbox}
\usepackage{subfig}
\usepackage{floatrow}
\usepackage{appendix}
\usepackage{url}

\DeclareSIUnit[number-unit-product = ]\percent{\char`\%}
\usepackage{hyperref}

\newcommand{\pal}{PAL}
\newcommand{\pmtwin}{2 $\mu$s}
\newcommand{\pmtextwin}{70 $\mu$s}
\newcommand{\mrdwin}{4 $\mu$s}
\newcommand{\beampeakE}{600 MeV}
\newcommand{\totalpot}{$8.26\times 10^{19}$}
\newcommand{\effdays}{107}
\newcommand{\partfile}{300 seconds}

\title{\boldmath First Light from Beam Neutrinos on an LAPPD in ANNIE}

\collaboration{\includegraphics[height=17mm]{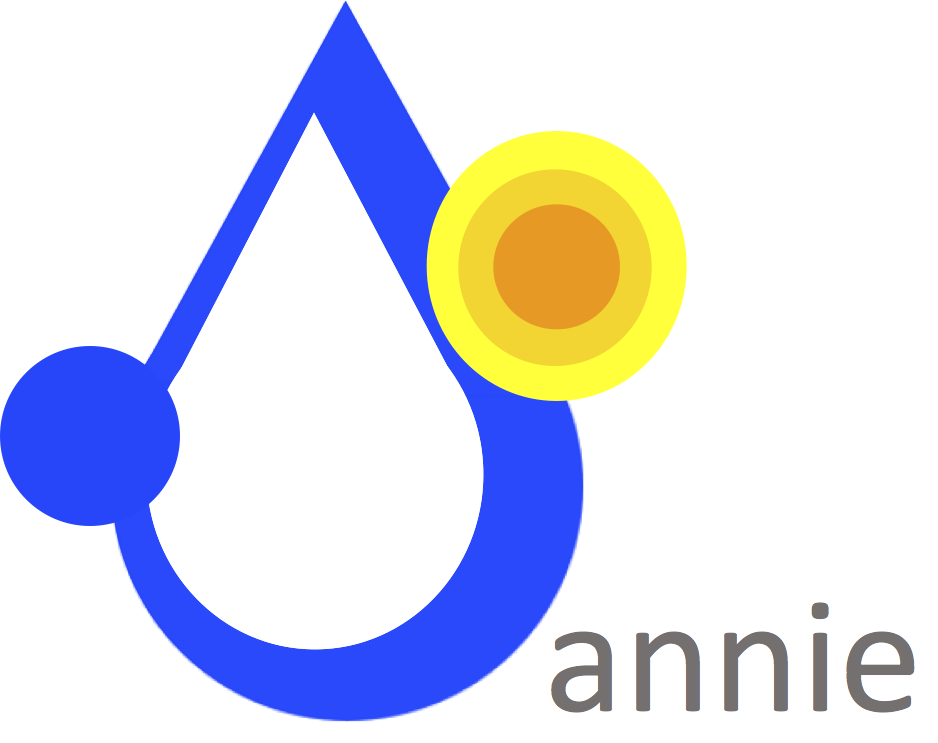}\\[6pt] The ANNIE collaboration}

\author[a]{S. Abubakar,}
\author[b]{B. W. Adams,}
\author[c]{D. Ajana,}
\author[c]{M. A. Aman,}
\author[d]{M. Ascencio-Sosa,}
\author[e]{A. Augusthy,}
\author[f,g]{Z. Bagdasarian,}
\author[h]{J. Beacom,}
\author[i]{M. Bergevin,}
\author[j]{D. Bick,}
\author[k]{M. Breisch,}
\author[d]{E. Brunner-Huber,}
\author[l]{G. Caceres Vera,}
\author[i]{S. Dazeley,}
\author[d] {S. Deng,}
\author[d] {S. Donnelly,}
\author[d]{S. Doran,}
\author[m]{E. Drakopoulou,}
\author[d]{S. Edayath,}
\author[n]{R. Edwards,}
\author[o]{J. Eisch,}
\author[d]{Y. Feng,}
\author[o]{V. Fischer,}
\author[p]{R. Foster,}
\author[o]{S. Gardiner,}
\author[e]{N. Goehlke,}
\author[q]{S. Gokhale}
\author[r]{A. Gupta,}
\author[l]{P. Hackspacher,}
\author[j]{C. Hagner,}
\author[l]{J. He,}
\author[k]{B. Kaiser,}
\author[a,s]{M. Kandemir,}
\author[d]{J. Kautz,}
\author[d]{F. Krennrich,}
\author[r]{M. Kumar,}
\author[k]{T. Lachenmaier,}
\author[t]{F. Lemmons,}
\author[d]{M. Lucas,}
\author[e]{D. Maksimovic,}
\author[p]{M. Malek,}
\author[e]{J. Martyn,}
\author[u]{A. Mastbaum,}
\author[o]{C. McGivern,}
\author[u]{J. Minock,}
\author[u]{C. Nguyen,}
\author[e]{M. Nieslony,}
\author[n]{M. O'Flaherty,}
\author[f,g]{G. D. Orebi Gann,}
\author[a]{B. K. Ozdemir,}
\author[i]{T. Pershing,}
\author[f,g]{L. Pickard,}
\author[r]{N. Poonthottathil,}
\author[d]{E. Pottebaum,}
\author[q]{C. Reyes,}
\author[n]{B. Richards,}
\author[q]{R. Rosero,}
\author[c]{M. C. Sanchez,}
\author[e]{D. T. Schmid,}
\author[v]{M. Smy,}
\author[d]{H. Sogarwal,}
\author[j]{M. Stender,}
\author[c]{A. Sutton,}
\author[l]{R. Svoboda,}
\author[d]{C. Sweeney,}
\author[a, w]{E. Tiras,}
\author[v]{M. Vagins,}
\author[d]{V. Veeraraghavan,}
\author[t]{J. Wang,}
\author[d]{A. Weinstein,}
\author[d,*]{M. Wetstein,\note[*]{Corresponding author.}}
\author[e]{M. Wurm,}
\author[q]{M. Yeh,}
\author[l]{T. Zhang}

\affiliation[a]{Erciyes University, Department of Physics, Kayseri, 38030, T\"urkiye}
\affiliation[b]{Quantum Optics Applied Research, Waldbronn 76337, Germany}
\affiliation[c]{Florida State University, Department of Physics, Tallahassee, FL 32306, U.S.A.}
\affiliation[d]{Iowa State University, Department of Physics and Astronomy, Ames, IA 50011, U.S.A.}
\affiliation[e]{Johannes Gutenberg Universit\"at, Institut f\"ur Physik \& EC PRISMA$^+$, Mainz 55128, Germany}
\affiliation[f]{University of California, Berkeley, Physics
Department, Berkeley, CA 94720 U.S.A.}
\affiliation[g]{Lawrence Berkeley National Laboratory, Nuclear
Science Division, Berkeley, CA 94720 U.S.A.}
\affiliation[h]{The Ohio State University, Department of Physics, Columbus, OH 43210, U.S.A.}
\affiliation[i]{Lawrence Livermore National Laboratory, Livermore, CA 94550, U.S.A.}
\affiliation[j]{Universit\"at Hamburg, Institut f\"ur Experimentalphysik, Hamburg 22761, Germany}
\affiliation[k]{Eberhard Karls Universit\"at, Kepler Center for Astro and Particle Physics, T\"ubingen 72076, Germany}
\affiliation[l]{University of California at Davis, Department of Physics and Astronomy, Davis, CA 95616, U.S.A.}
\affiliation[m]{N.C.S.R. "Demokritos", Institute of Nuclear and Particle Physics, Agia Paraskevi 15341, Greece}
\affiliation[n]{University of Warwick, Department of Physics, Coventry CV4 7AL, U.K.}
\affiliation[o]{Fermi National Accelerator Laboratory, Batavia, IL 60510, U.S.A.}

\affiliation[p]{University of Sheffield, Department of Physics and Astronomy, Sheffield, S10 2TN, U.K.}

\affiliation[q]{Brookhaven National Laboratory, Upton, NY 11973,
U.S.A.}
\affiliation[r]{Indian Institute of Technology Kanpur, Department of Physics, Kanpur 208016, India}
\affiliation[s]{Recep Tayyip Erdogan University, Department of Physics, Rize, 53100, Türkiye}
\affiliation[t]{South Dakota School of Mines and Technology, Physics Department,  Rapid City SD, 57701, U.S.A.}

\affiliation[u]{Rutgers University, Department of Physics and Astronomy, Piscataway, NJ 08854, U.S.A.}
\affiliation[v]{University of California at Irvine, Department of Physics and Astronomy, Irvine CA, 92697, U.S.A.}
\affiliation[w]{University of Iowa, Department of Physics and Astronomy, Iowa City, IA 52242, U.S.A.}

\emailAdd{wetstein@iastate.edu}

\abstract{The Accelerator Neutrino Neutron Interaction Experiment (ANNIE) is both a physics experiment and a technology testbed for next-generation light-based neutrino detection. In this paper, we report the first demonstration of a fully integrated Large Area Picosecond Photodetector (LAPPD) operating in a running neutrino beam experiment. Particular focus is given to the design, commissioning, and successful deployment of the Packaged ANNIE LAPPD (PAL), a waterproof, self-triggering module incorporating fast waveform digitization and precision timing synchronized to the ANNIE detector subsystems. We identify beam-correlated LAPPD data frames consistent with charged-current neutrino interactions observed in multiple detector subsystems, establishing the first detection of neutrino-induced Cherenkov light with an LAPPD. These results validate the system-level performance of LAPPDs under realistic experimental conditions—including long-term stability, timing synchronization, and event matching with conventional PMT and muon detector systems—marking a critical step toward their deployment in future large-scale neutrino and particle detectors.}


\begin{document}
\maketitle
\flushbottom

\section{Introduction}
\label{sec:intro}

LAPPDs are MicroChannel Plate (MCP)-based~\cite{mcpref} imaging photosensors with a variety of potential applications in high-energy and nuclear physics, as well as practical applications in medical imaging~\cite{LAPPDhist} and other areas. Produced using advanced fabrication techniques such as atomic layer deposition, LAPPDs offer ${\sim}10^6$ gain~\cite{IncomPerformance}, approximately 50 ps time resolution~\cite{timingpaper}, and unprecedented large area coverage. At 20 cm x 20 cm, they have a form factor nine times larger than the next-largest MCP-PMTs, at significantly lower cost per unit area. The large LAPPD area is coupled with a unique ability to correlate photon arrival times with sub-cm position resolution, determined primarily by the structure of the anode. 

Until this point, studies of LAPPDs have been limited to proof-of-principle studies on the bench-top, focused primarily on characterizing LAPPD response in the single-photoelectron regime. Investigating LAPPD performance in the context of a running experiment, using actual neutrino interactions, represents a critical milestone toward establishing LAPPDs as viable next-generation photodetector for particle physics applications. Here we present the first neutrino interactions detected using an LAPPD in a working experiment and document the relevant technical details of the system developed to further the application readiness of this technology. In Section~\ref{sec:ExperimentalContext}, we briefly describe the ANNIE detector and the experimental context. In Sec. \ref{sec:Technical}, we describe the integration of an LAPPD into the ANNIE experiment. Finally, in Section~\ref{sec:results1}, we establish the first detection of beam neutrinos with an LAPPD.

\section{Review of the ANNIE Experiment}
\label{sec:ExperimentalContext}

ANNIE is both a physics experiment studying neutrino-nucleus interactions in a water target and a unique US platform for developing next-generation neutrino detection technology~\cite{ANNIEProposal}. The primary physics goal of ANNIE is to study neutrino interactions on water, with specific focus on measuring the abundance of final state neutrons from neutrino-nucleus interactions, using gadolinium (Gd) enhanced water. Measurements of final-state neutron multiplicity will improve our understanding of neutrino-nucleus interactions. The complex many-body physics of neutrino-nucleus scattering is among the dominant systematic uncertainties in precision long-baseline oscillation measurements~\cite{{oscillationsE},{oscillationsE2}}. 

ANNIE also supports a broad R\&D program that serves the experiment's core physics goals while developing and testing new technologies for the future. One of the central goals of ANNIE is the first demonstration of LAPPDs in a neutrino experiment, as described in this paper. Other milestones of the R\&D program include deployment of the first Gd-loaded water target in a neutrino beam (2021) and the first water-based liquid Scintillator (WbLS) target in a neutrino beam~\cite{ANNIE:2023yny}. 

\subsection{The ANNIE Detector}

\begin{wrapfigure}{r}{3.6in}
 \vspace*{-0.2in}
  \includegraphics[width=3.4in, angle=0.0]{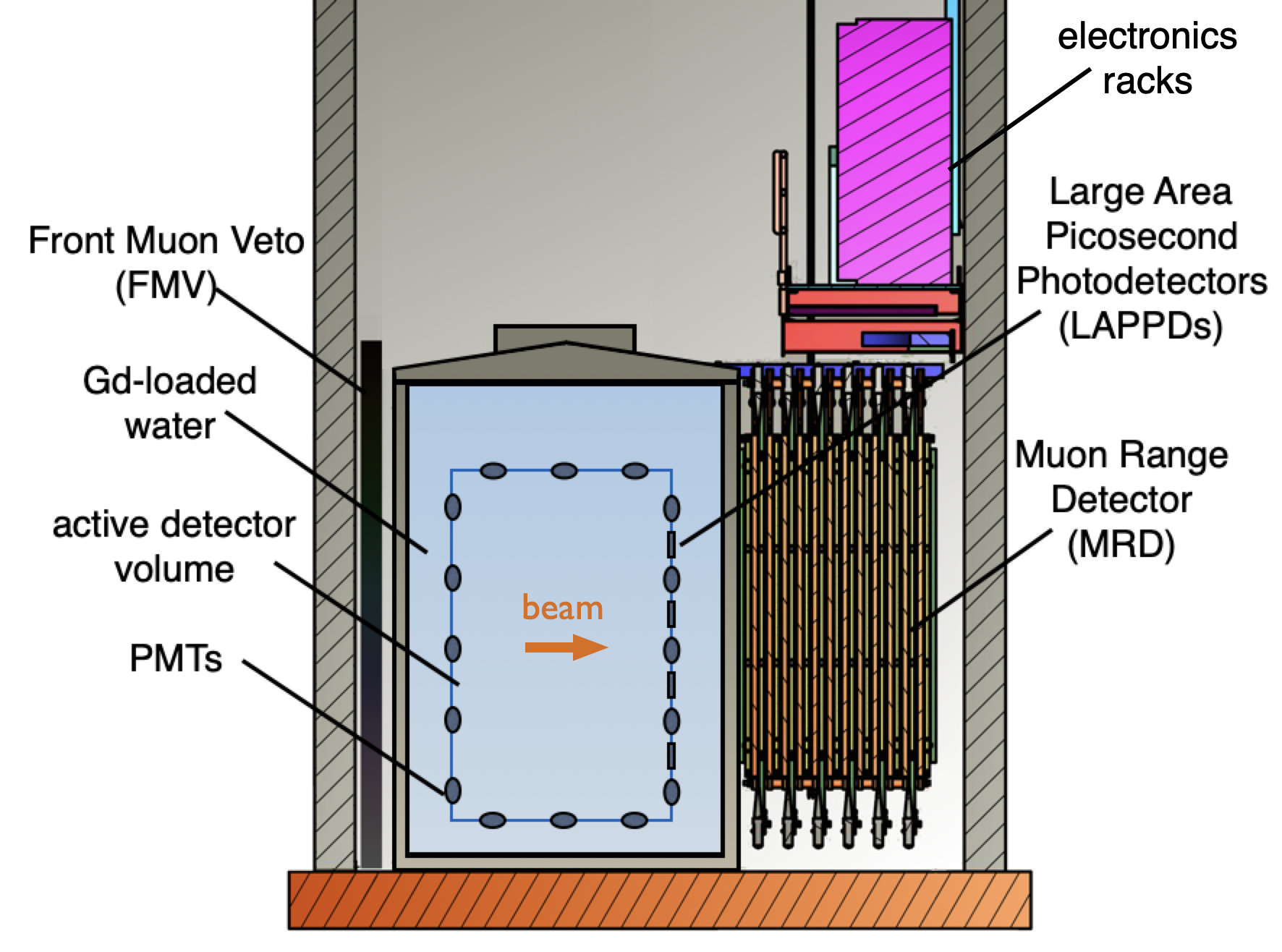}
 \vspace*{-0.1259in}
 \caption{The ANNIE Detector.}
\label{ANNIEdetector}
\end{wrapfigure}

Figure~\ref{ANNIEdetector} shows the ANNIE detector configuration. The main target consists of an upright cylindrical steel tank filled with 26 tons of Gd-loaded (0.1\% by mass) ultra-pure deionized water. The photodetectors inside the tank are a mix of 8-, 10- and 11-inch conventional photomultiplier tubes (PMTs) that provide coverage in all directions. Eight regularly spaced access ports around the lid of the water tank each permit a column of up to three LAPPDs to be inserted into the tank. Data presented in this paper was acquired with a single LAPPD at a central position on wall of the tank downstream of the beam.
In addition to the water volume, the full ANNIE detector system also includes a downstream muon detection system (refurbished from SciBooNE's Muon Range Detector, or MRD~\cite{SciBooNE:2008bzb}) that provides the primary constraint on muon energy and direction, as well as a scintillator-based Front Muon Veto (FMV) upstream of the water target. 

A muon produced by a neutrino interaction in the fiducial volume will propagate through the tank, producing Cherenkov light along its path, before entering into and passing through the layers of scintillator paddles and iron absorber in the MRD. The position and timing of scintillator hits, combined with information about the Cherenkov light distribution from the tank photodetectors, are used to reconstruct the muon track. Neutrons produced by the neutrino interaction scatter and lose energy (thermalize) allowing them to be captured on either H or Gd in the active volume. The Gd loading substantially increases the neutron detection efficiency by increasing the capture cross-section for thermal neutrons (thereby shortening the neutron capture time) and providing a higher-energy, more easily-detected $\sim$8 GeV de-excitation $\gamma$-ray cascade~\cite{FSneutrons,EGADS}.

Adding the LAPPD data to the information provided by conventional photodetectors and the MRD allows ANNIE to achieve its target precision for both muon track reconstruction and the localization of the neutrino interaction vertex for charged-current interactions. 
In order to ensure the best possible containment of neutrons produced by neutrino-nucleus collisions, the data sample for ANNIE neutron multiplicity measurements must be restricted to neutrino interactions taking place inside a narrow fiducial volume within the tank volume. Reducing the uncertainty in the interaction vertex improves the accuracy---and therefore the effectiveness---of this fiducialization. LAPPDs therefore play an important role in ANNIE's physics program, particularly the neutron multiplicity measurement.

\subsection{Booster Neutrino Beam}

The ANNIE detector and data acquisition system are designed around the Booster Neutrino Beam (BNB)~\cite{MiniBooNE:2008hfu}, which delivers neutrinos in discrete bursts timed to the accelerator Radio Frequency (RF) structure. The low duty cycle of the BNB allows the experiment to statistically suppress continuum backgrounds by confining the readout of the various detector sub-systems to a short time window around the arrival of neutrinos. A subset of off-beam data provides a sample of the continuum backgrounds, while laser, LED, and radioactive sources are used for calibration.

BNB neutrinos are produced from the decay of mesons made by colliding 8 GeV protons with a stationary beryllium target located at the MI-12 facility. Pions in a narrow range of momenta are selected using magnetic horns and steered to a decay pipe, where they decay to form a beam of 93$\%$ pure muon neutrinos (with 6$\%$ muon anti-neutrino and 1$\%$ electron (anti) neutrinos) with an energy spectrum peaked at a little below \beampeakE\/~\cite{MiniBooNE:2008hfu}.

The Booster Neutrino Beam (BNB) is extracted from the Booster accelerator complex. Protons are organized and accelerated by the Booster in radiofrequency (RF) buckets, stable regions in phase space where beam can be captured and accelerated. A bucket filled by captured beam is called a bunch. The Booster RF sweeps from 37.8 MHz at injection to 52.8 MHz at extraction. The resulting 1.6 $\mu$s-long series of bunches, known as a spill, is delivered on the 15 Hz frequency of the linear accelerator (Linac) at the start of the accelerator complex. Because proton batches are shared with other beamlines, ANNIE receives an average spill rate of roughly 5 Hz. 

\subsection{System Timing and DAQ Software}
\label{sec:systtiming_daq}

\begin{figure}[]
	\begin{center}
		\includegraphics[width=0.85 \linewidth]{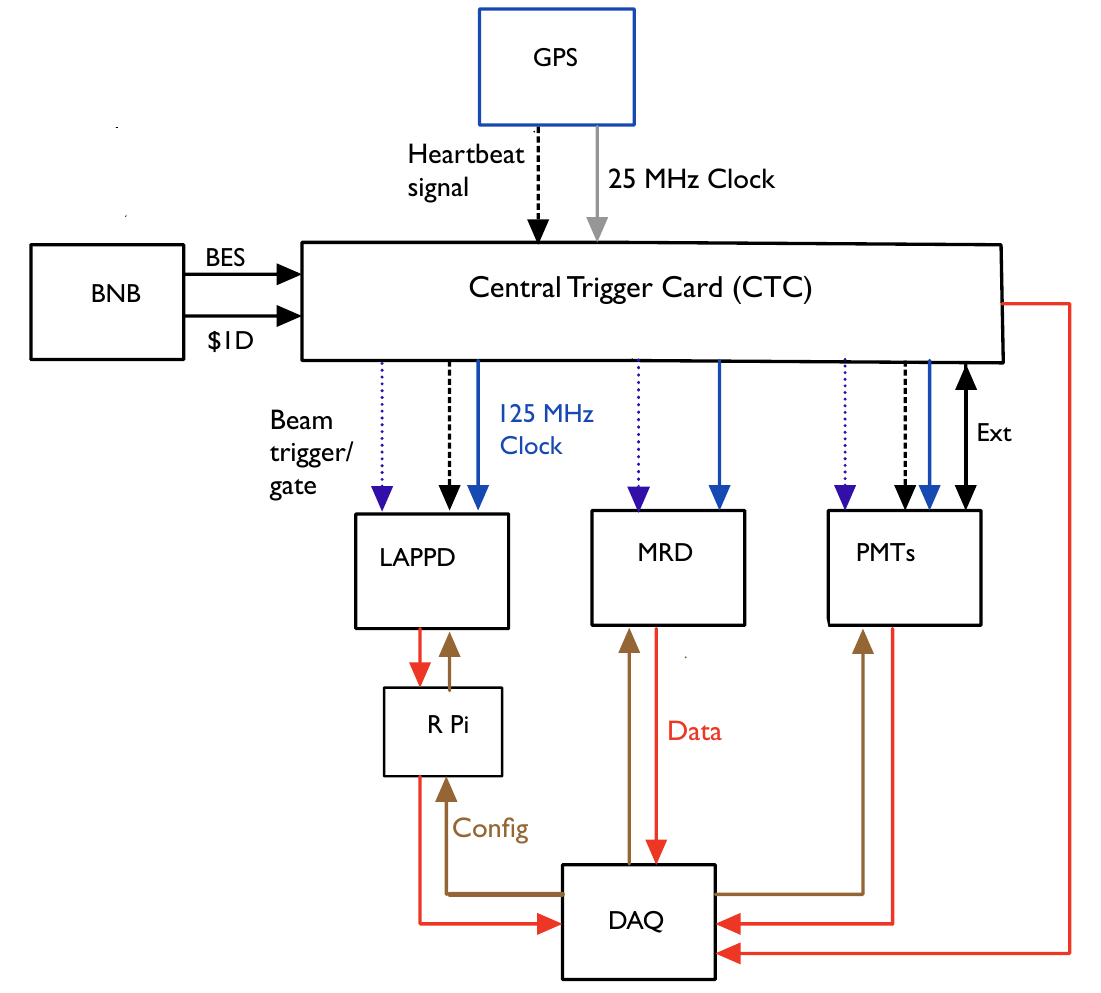}
	\end{center}
	\caption{Diagram illustrating the integration of the LAPPD subsystem into the overall timing, trigger, and data flow for ANNIE. Solid red arrows show the path of data flow and solid brown arrows the path of configuration information. The 25 MHz clock signal from the GPS is shown as a solid gray arrow. All copies of the heartbeat signal are shown as dashed arrows, copies of the 125 MHz clock signal from the CTC are shown in solid blue, and copies of the beam trigger (treated as a gate by the LAPPD subsystem and as a forced readout trigger by all other systems) are shown as dotted blue arrows. The black bidirectional arrow labeled ``Ext'' indicates two-way communication between the CTC and PMT readout electronics, needed to extend the nominal readout window for neutrino interactions likely to contain neutrons.}
		\label{fig:daq_integration}
\end{figure}

Figure \ref{fig:daq_integration} provides a visual summary of the flow of timing and trigger information between the different components of the ANNIE experiment. 

The system clock for the entire ANNIE experiment is provided by a CAEN V1495 general purpose VME board with a customizable FPGA, dubbed the ANNIE Central Trigger Card (CTC) \cite{V1495}. The CTC is disciplined to a 25 MHz clock signal provided by a customized Spectrum Instruments TM-4 GPS clock with a double-ovenized crystal oscillator \cite{TM4}. The CTC multiplies up the 25 MHz input clock to 125 MHz and passes that clock to the various subsystems.  
All subsystems match their timing to this clock via phase-locked loops (PLLs). Where needed, the CTC forwards a  heartbeat, or pulse-per-second (PPS) signal, also provided by the GPS clock, to the various subsystems. 

The CTC also manages, forwards, and timestamps a variety of beam-related and calibration triggers. Two of these signals are of particular importance. 
The CTC forms a \emph{beam trigger} or \emph{beam gate} as the logical and of a fiber-optic copy of the Booster Extraction Sync (BES) signal which precedes beam extraction by 320 $\mu$s, and the Tevatron clock \$1D signal, which indicates that the beam will be accelerated and extracted to the BNB and hence delivered to ANNIE. It then forwards time-delayed copies of this \emph{beam trigger} to the readout electronics for the conventional PMT readout system and MRD, triggering both systems to read out all data related to a spill. 

The CTC forwards an undelayed copy of the \emph{beam trigger} to the LAPPD electronics, which apply their own internal delay. The use of the beam trigger by the LAPPD electronics also differs from the other subsystems. LAPPD readout electronics digitize signals at very high time resolution and consequently operate with a digitization buffer much shorter than the typical length of a BNB spill. In order to minimize deadtime and maximize efficiency in the window of interest, the LAPPDs are therefore designed to trigger autonomously within a configurable time-window (currently $20 \thinspace {\mu}\mathrm{s}$) after receipt of a beam trigger, usually referred to in this context as a \emph{beam gate}.  

For DAQ software, ANNIE uses the ToolDAQ framework, developed at Warwick University for the HyperK collaboration~\cite{benajmin_richards_2018_1482767}. Asynchronous modular code sequences (ToolChains) pull data from each of the subsystems. 
An overall system loop on the top-level DAQ computer queries and pulls data from each subsystem ToolChain when available. CTC data, which includes timestamps for the beam input and heartbeat signals, is also pulled as a data stream. These time stamps are used to tie together the responses from individual subsystems to build a complete record of the detector's response to a beam spill, cosmic ray, or calibration source, hereafter referred to as an event.

\section{Technical Integration of an LAPPD into ANNIE}
\label{sec:Technical}

{ 
The Packaged ANNIE LAPPDs ({\pal}s) are waterproof modules containing the LAPPD along with all the electronics necessary to digitize and read out the LAPPD signals and monitor the environment within the housing. Electronics on the surface provide a control and monitoring interface to the {\pal}s, manage data flow between the {\pal}s and DAQ, and forward signals from the CTC to the {\pal}s.
The two sub-systems are physically connected by waterproof cables, described in greater detail in Appendix \ref{app:AppC}. 

The design of this system was shaped by several key design requirements. Fast sampling electronics and synchronization of the LAPPDs and other subsystems with a reliable central clock are necessary for ANNIE’s ultimate goal of demonstrating 100ps global time stamping of photon hits. Power dissipation from the readout electronics needed to be minimized to reduce heating effects at the LAPPD and in the interior of the waterproof housing.  This ensured stable operating conditions (the resistance and performance of the MCPs can be temperature-sensitive) and also minimized temperature-related physical stresses on the LAPPD. Signals and power were routed between the surface and the tank in a way that maximized signal fidelity and power stability while minimizing the risk of water leaks.}

\subsection{Surface Electronics }

Several components of the ANNIE LAPPD subsystem are located in standard racks on the second level of the experimental hall, as shown in Figure~\ref{ANNIEdetector}. These ``surface electronics'' components provide synchronization and clock signals, coordinate data transfer and concentration, pass data from the LAPPD system on to the DAQ computer, supply system power, and monitor system conditions.
The components that forward power from the power bus, provide slow controls, and monitor conditions inside the \pal\/ through the high-density waterproof cables are discussed in detail in Appendices \ref{app:AppC} and  \ref{app:AppA}.

VME-based Annie Central Cards (ACCs) provide communications and clock synchronization with the underwater readout system, and serve as conduits for transferring the digital outputs of the LAPPDs to the ANNIE DAQ computer using a Raspberry Pi. Each ACC card is capable of servicing up to 2 LAPPDs. The ACCs are controlled by Altera Arria V FPGAs~\cite{ArriaV}. The communication link connecting the ACC and ACDC boards relies on the Low Voltage Differential Signaling (LVDS) logic standard. Each link between an ACC and ACDC consists of 4 twisted pairs: two lines for transmitting and receiving data, one line for sending the beam gate, and a clock signal for ensuring clock-domain alignment. 
A USB-serial buffer chip operating at 48 Mbps transfers data between the ACC and the control computer. While the ACC cards are designed to sit in a VME crate and have a standard VME form-factor, they do not require a VME backplane. They are housed in a VME crate with no backplane and powered by an Excelsys power supply\cite{Excelsys}. 

Each ACC is synchronized to the GPS-disciplined CTC 125 MHz clock via a PLL and forwards copies of this clock signal to the {\pal}s. The ACC delays the \emph{beam gate} signal by a configurable amount, currently set to 325.25 microseconds.
Because there were a limited number of data lines in the waterproof cables, the ACC multiplexes the delayed beam gate signal with the heartbeat signal from the CTC before sending it to the \pal\/.  In order to limit the number of synchronization frames, the ACC applies a configurable prescale to the 1 PPS heartbeat. 

\subsection{Packaged ANNIE LAPPD (\pal)}
\label{subsec:waterproof-package}

\begin{figure}[]
	\begin{center}
		\includegraphics[width=1.0 \linewidth]{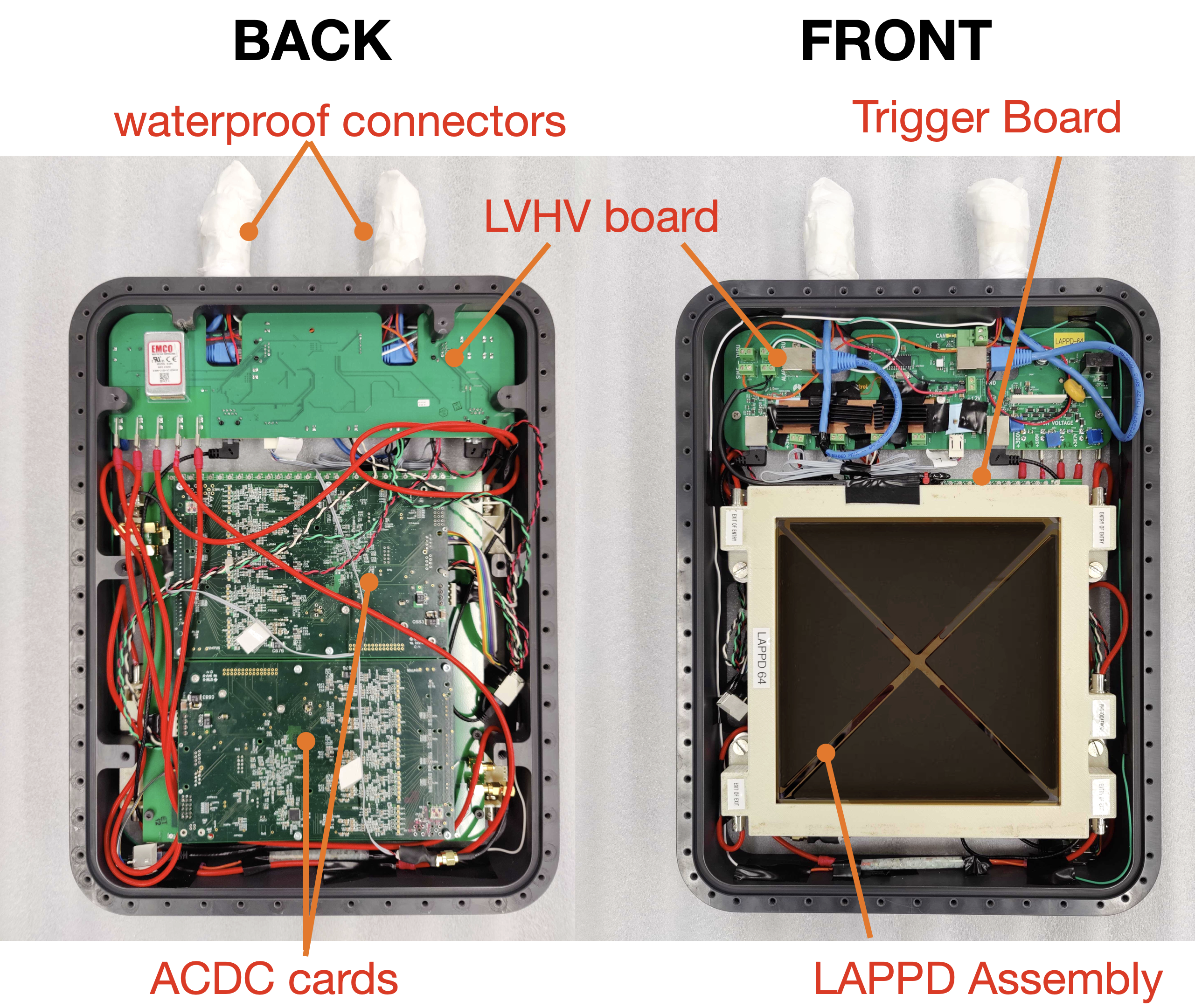}
	\end{center}
	\caption{Photographs of the back (left) and front (right) of a \pal\/. The module is oriented in the direction it is deployed in the water, with the waterproof connectors pointing upward. At the top of the modules is the LVHV board, which manages power and slow controls, and routes the communications line. The LAPPD assembly consists of the LAPPD mounted onto an analog pickup board with two bridge circuits: a powered trigger card (top) and passive bridge circuit (bottom). Plugged into the back of the LAPPD assembly are the two ACDC readout cards.}
		\label{fig:paldiagram}
\end{figure}

A \pal\/ consists of a waterproof housing containing the LAPPD assembly with readout electronics (ACDC cards), the slow controls board (LVHV), and independent leak-sensing and temperature-monitoring devices, as well as flood-proof connectors that mate to waterproof cables. These elements are pictured in Figure~\ref{fig:paldiagram}. 

\begin{figure}[]
	\begin{center}
		\includegraphics[width=1.0 \linewidth]{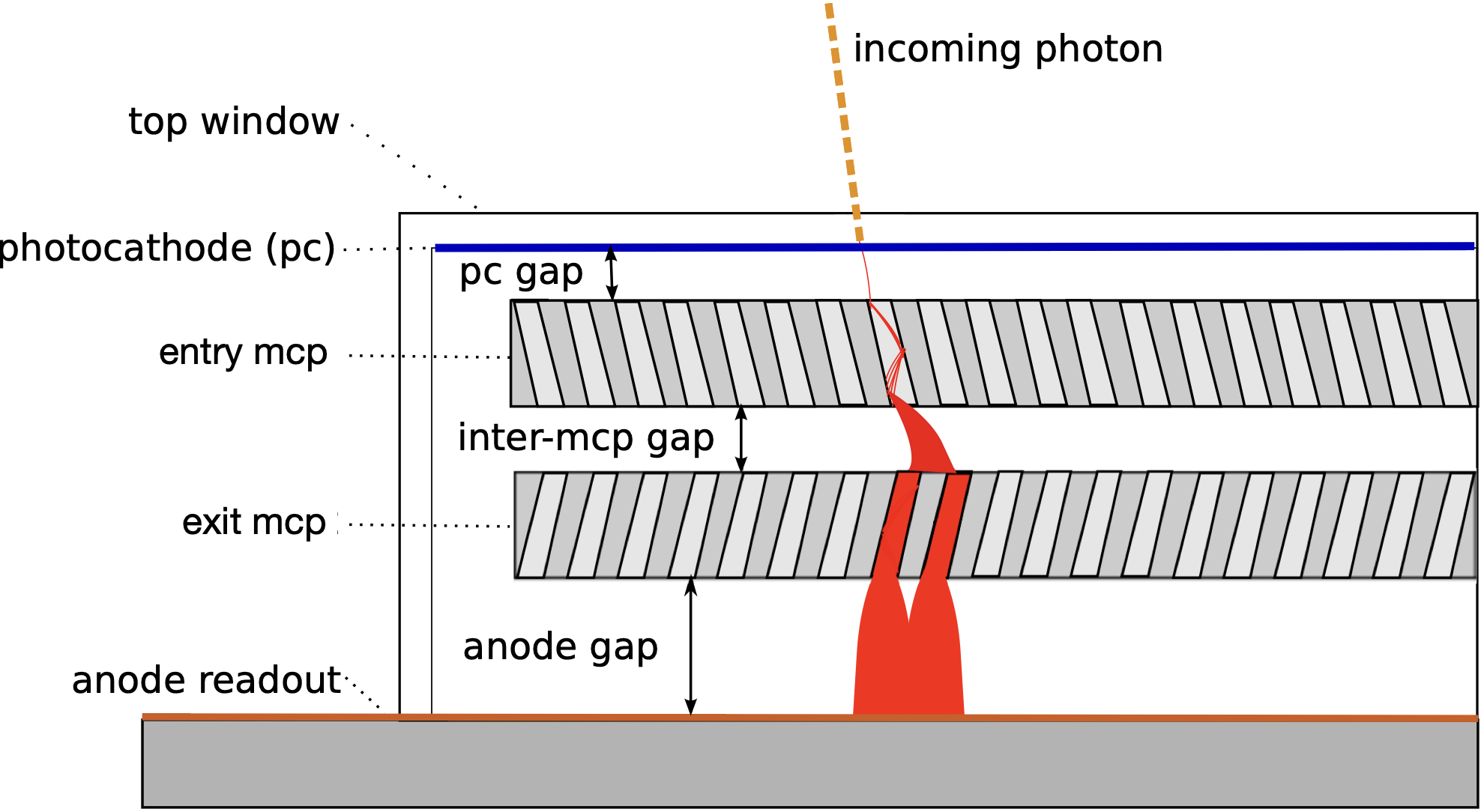}
	\end{center}
	\caption{A cross-section showing the response of an LAPPD to a single photon.}
		\label{fig:lappdschematic}
\end{figure}

Figure~\ref{fig:lappdschematic} illustrates a functional cross-section of an LAPPD in operation. Light is incident on a photocathode producing a photoelectron (PE). The PE produced at the photocathode is accelerated into a porous MCP gain stage (the Entry-MCP).  Inside the 20-micron pores, the electron collides producing an avalanche of secondary electrons. These secondary electrons are then accelerated across a gap, into a second MCP (the Exit-MCP). Collisions within the second MCP produce an avalanche, further amplifying the signal. The cloud of electrons ultimately drifts to the anode. 

The first LAPPD deployed in ANNIE, LAPPD-40 (numbered reflecting its production order by the manufacturer, Incom Inc.), is a first-generation model LAPPD~\cite{IncomPerformance} instrumented with 28 silk-screened silver anode microstrips that are integral with the module. Photon hits are reconstructed based on the arrival of signal pulses at the two ends of each microstrip, as detailed further in Ref~\cite{multiphoton}. The nominal gain of LAPPD-40, as measured by Incom, is $1.6 \times 10^{6}$ at 950V per MCP, and the average Quantum Efficiency (QE) is $\sim$17$\%$. This performance is on the low end of the LAPPDs received by ANNIE, due in part to LAPPD-40 being one of the earlier functioning models.

\subsubsection{Structure of the \pal}

In order to fully exploit the native timing resolution of the LAPPD, signals must be digitized close to the detector. This is one of the primary design drivers for this system, as it requires us to locate the readout boards in the water, contained within a waterproof housing. Doing so minimizes the number of independent signals that need to be sent from the LAPPD assembly to the surface. This minimizes both the number and width of the penetrations in the housing.

\begin{figure}[]
	\begin{center}
		\includegraphics[width=1.0 \linewidth]{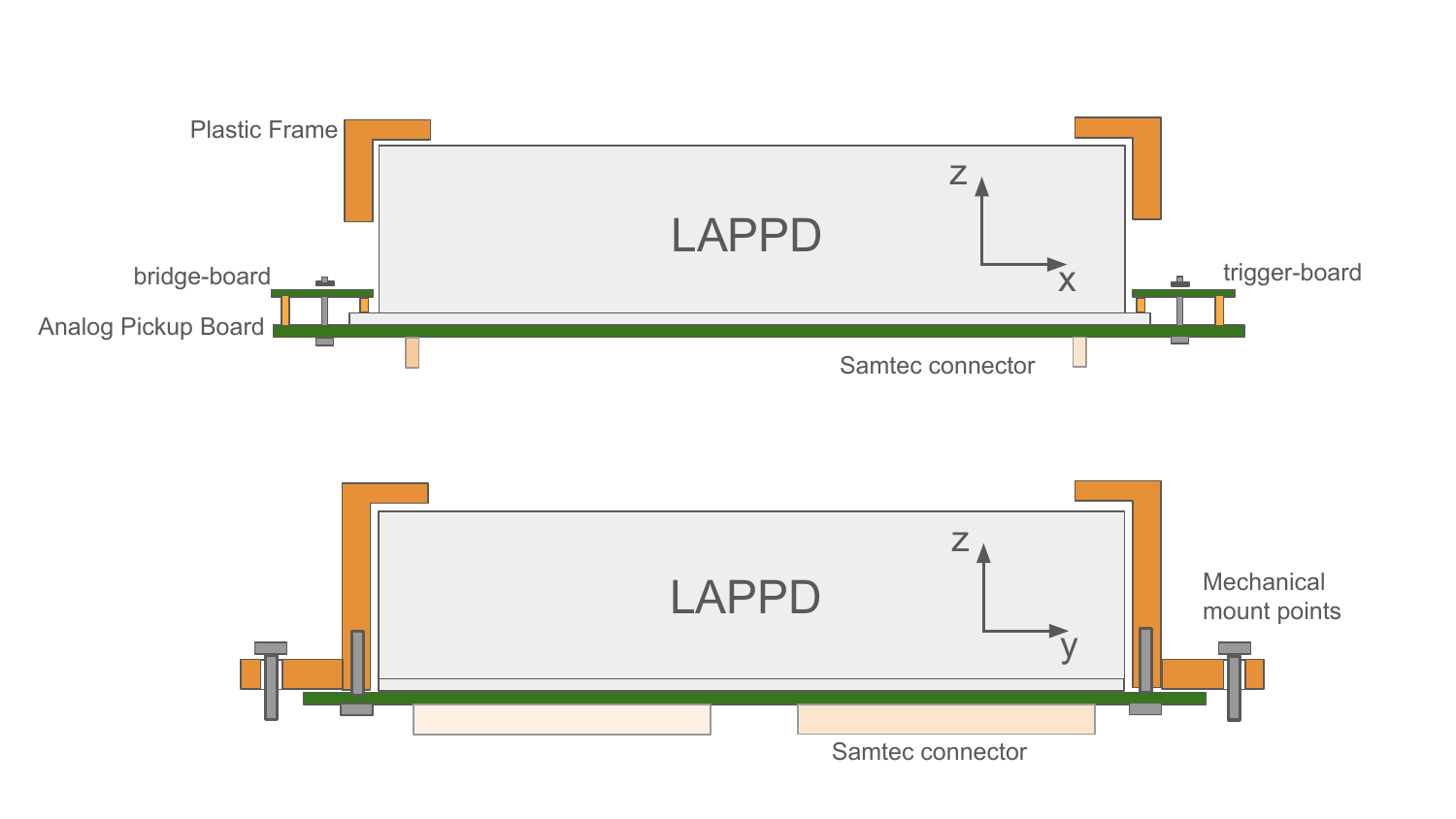}
	\end{center}
	\caption{Two views of the LAPPD frame, emphasizing the mechanical elements. The only elements that directly contact the LAPPD packaging are the pogo pins of the bridge- and trigger-boards, shown in the upper panel, and the high-voltage pogo pins, which are not shown. The plastic LAPPD frame does not directly touch the LAPPD.}
		\label{fig:lappdassembly}
\end{figure}

Figure~\ref{fig:lappdassembly} shows a side-on view of the LAPPD assembly. The glass tile is encased in an Ultem frame with mounting tabs and mounted onto a custom ANNIE Analog Pickup Board (APB), which brings the signals to connectors that can interface with the readout electronics. The microstrips on the glass bottom-plate extend out of the vacuum region of the detector onto a lip that extends 4.6 mm out on either side of the LAPPD. Two {\it bridge boards} establish contact via spring-loaded pogo pins between the silver microstrips of the LAPPD and the APB. One of these bridge boards is passive and provided by the manufacturer, the other is a custom-designed active trigger-board that passes the signals on to the pickup board but also senses pulses above a set threshold (see Section~\ref{subsec:readout-trigger}). 

The APB routes the signals from both sides of the 28 anode microstrips to two high-density Samtec connectors facing away from the back of the package. Two 30-channel ACDC cards plug into these high-density connectors. Each ACDC card reads both sides of the microstrips for half of the LAPPD, which ensures that differential timing calculations between the two sides of the microstrip are handled on the same card. The traces of the analog pickup board are designed to follow approximately equal lengths to each of the chips, in order to minimize channel-to-channel timing skew.

SHV cables attached to the Ultem frame and terminated with spade-connectors plug into a board that provides power and slow controls, as well as an AC-coupled pass through of critical signals sent from the surface into the water. Since this board provides both low-voltage power to the readout electronics and high-voltage to the various stages of the LAPPD, it has been dubbed the low-voltage high-voltage (LVHV) board (see Appendix~\ref{app:AppA}). The mechanical aspects of the LAPPD assembly are diagrammed in Figure~\ref{fig:lappdassembly}.

The physical waterproof housing of the \pal\/ consists of a 2-inch thick PVC sidewall sandwiched between a stainless steel backplate and a quarter-inch thick UVT acrylic window. The window and backplate are bolted onto the sidewall with
chemical-resistant Viton fluoroelastomer 3/16-inch O-rings~\cite{Viton} providing the waterproof seal. All of the power and communications lines connect to the LVHV board, which is affixed to the inside top of the sidewall. The LAPPD assembly is mounted onto tabs extending from the sidewall. All HV cables, power and communications lines from the ACDC cards and trigger cards plug into the LVHV board.

A salt bridge, consisting of a punctured straw filled with road salt and sandwiched between two terminals, is connected to a pair of leads that connect to power/control lines on one of the waterproof cables. The salt bridge serves as a catastrophic leak detector.  A second pair of power/control lines connect to a thermistor that is placed at the midpoint of the housing, close to the LAPPD itself.

\subsubsection{Digitization and Triggering in the \pal}
\label{subsec:readout-trigger}

We required cost-effective, low-power readout electronics that fit within a small form factor of the waterproof housing, and could be easily integrated with slow controls. 

LAPPDs provide signals in the form of negative-going pulses at amplitudes of tens of mV and durations of about 1 ns. In order to fully exploit the rich timing information provided by these photodetectors, we rely on high frequency waveform-sampling electronics that acquire these signals continuously at a rate of 10 Gigasample per second (GS/s) and store them in analog front-end buffers. When a trigger condition is met, acquisition is stopped, and the samples are digitized. The latter process takes a considerable amount of time, typically several microseconds, and should be performed only at need. The decision when to digitize must be made by trigger circuitry, which may be built into the waveform samplers, or may be external.

Sampling and digitization of LAPPD waveforms is handled by ACDC cards, custom boards designed at the University of Chicago, that attach to the APB through high-density connectors. The ACDCs are built around the 6-channel PSEC4 waveform sampling ASIC, which provides 10 GS/s sampling rates and 1 GHz analog bandwidth~\cite{PSEC}. 
Each ACDC card contains five PSEC4 chips, for a total of thirty channels of readout. Two ACDC boards are used for each LAPPD, to handle the full 56 channels of the photosensor, with 2 spare channels on each board. Signals can be coupled to the spare channels through SMA connectors on the pickup board.  

Ovenized oscillators onboard the ACDCs are disciplined via PLL to the 125 MHz clock sent by the ACC.
An onboard FPGA on each ACDC runs at 320 MHz, multiplied up from the 40 MHz clock provided by the oscillator, and handles communications, internal logic, and timestamping of all received signals. The 256-sample ring buffers of the PSEC4 chips on the ACDCs are also latched onto the 40 MHz from the GPS-disciplined oscillator.
 
 The LAPPD readout electronics were bench-tested to verify that channel-to-channel timing alignment was better than the 100 ps sampling interval. Both in bench tests and in situ, a loss of synchronization between PSEC4 chips would manifest as a simultaneous 8 ns shift across five readout channels; no such occurrences were observed. Calibration with synthetic signals was used to measure and correct path-length differences in the PCB pickup board traces. Small biases of order 10 ps may remain, arising from non-uniform PSEC4 sampling rates, and will be investigated in future work. These nonuniformities fall well below our current target of 100 ps single-photon time resolution.

The built-in trigger circuitry of this version of the ACDC (Rev. B) was not sufficiently sensitive to capture low-amplitude pulses for the ANNIE use case. In order to improve the trigger sensitivity and to provide multi-channel coincidence logic, a dedicated trigger circuit was designed as a drop-in replacement for one of the passive, manufacturer-provided bridge boards that connect the LAPPD anode to the APB (see Figure~\ref{fig:lappdassembly}). The requirement that the trigger circuit be a drop-in replacement meant that the electronics monitoring the 28 individual signal channels, as well as the actual multiplicity logic, had to fit within a small form factor. Further, the power dissipation of the entire trigger circuit had to be limited to 1W or less and noise pickup and crosstalk had to be minimized. This is accomplished by using low-voltage differential-signaling (LVDS) internally to the circuit. 

The ACDC separates the multiplexed beam gate and heartbeat signals on the basis of their distinct signal widths. Data from any LAPPD self-trigger that occurs within the configurable \emph{beam gate} width (currently set to 20 $\mu$s) is recorded. The ACDCs record timestamps for all received beam gate signals and all self-triggers. Short synchronization (sync) data frames record the local time stamp of the heartbeat signal. For most of the data-taking period, one in ten heartbeat signals were recorded (i.e. one sync frame every ten seconds). 
 
\subsection{LAPPD Deployment}

The ANNIE detector is designed to enable easy deployment and removal of the {\pal}s \emph{ in situ} to the filled water tank. Columns for LAPPD deployment are located at the 8 vertices of the octagonal inner structure of the tank. Each of the 8 steel legs of the inner structure contains a Unistrut track. {\pal}s mount onto PVC panels (dubbed \emph{surfboards}), which slide down into these tracks through rectangular openings (deployment ports) on the top of the tank. Between one and three {\pal}s
can be accommodated on a single surfboard.  

\begin{figure}[]
	\begin{center}
		\includegraphics[width=1.0 \linewidth]{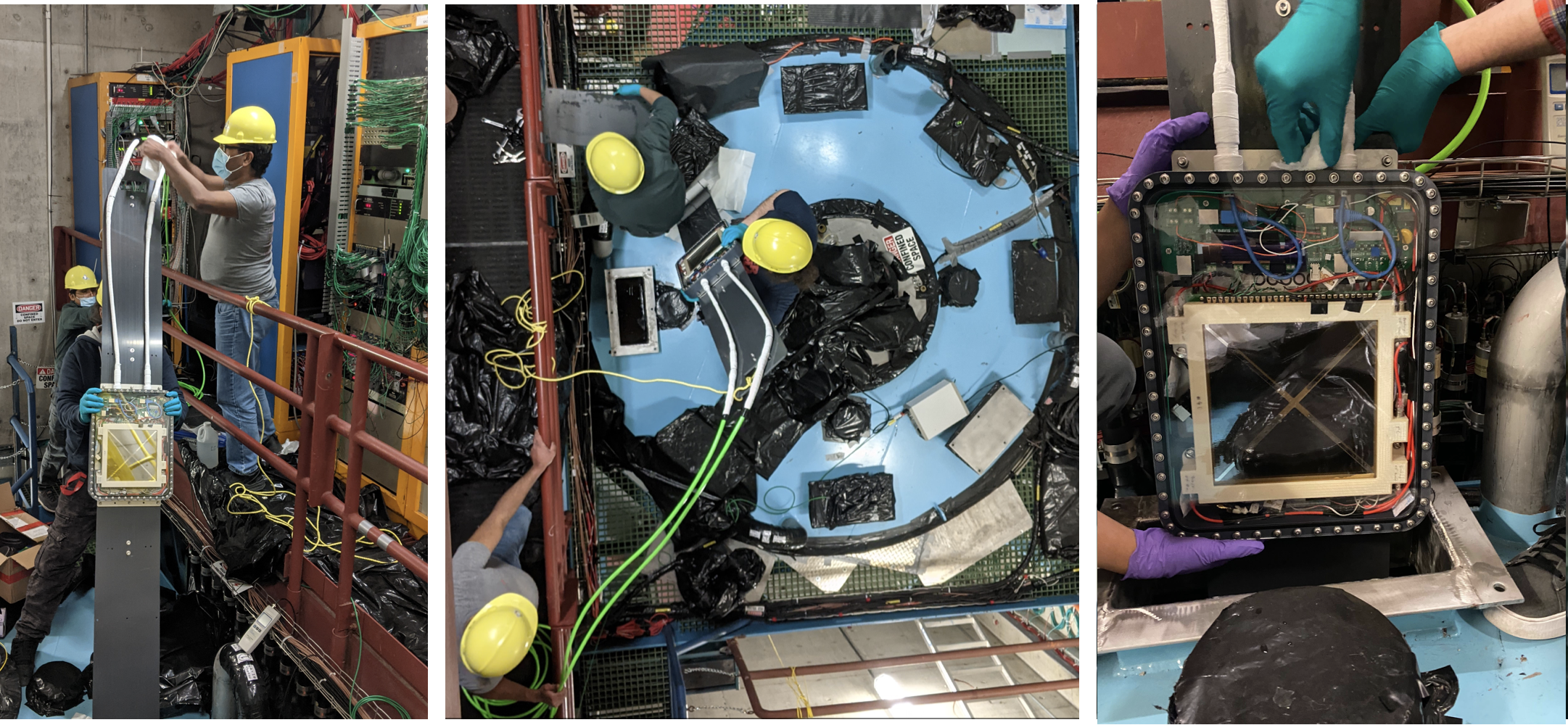}
	\end{center}
	\caption{Photographs from the deployment of LAPPD-40 in March 29, 2022: Preparing the surfboard (left), overhead view of the surfboard being lifted toward a open LAPPD deployment port (center), and lowering the surfboard into the port (right).}
		\label{fig:deployment_triptych}
\end{figure}

The first ANNIE LAPPD was deployed on March 29, 2022 and integrated with the full DAQ by April. Figure~\ref{fig:deployment_triptych} shows photographs of the preparation and deployment of the surfboard with LAPPD-40 mounted.

LAPPDs can be distributed isotropically in the azimuthal and $z$ directions, but those in the forward direction, downstream of the beam, are the most critical for the purposes of reconstructing the Cherenkov light from forward-going muons. The first LAPPD was therefore deployed at the most central position possible downstream of the beam. 

\section{Observation of First Beam Neutrinos with an LAPPD}
\label{sec:results1}

We report here studies performed with data taken between June 2022 and July 2023. During this period, there were a total of \effdays\/ days where the beam was stable, all ANNIE subdetectors were functioning nominally, and the LAPPD was fully operational and integrated with the ANNIE DAQ. The data taken during this \effdays-day period corresponds to an integrated exposure of \totalpot\/ protons-on-target (POT). 

\subsection{Event Building}

The ANNIE DAQ receives and stores information from the different subsystems (central trigger, MRD, tank PMTs, LAPPDs) in parallel streams as described in Section~\ref{sec:systtiming_daq}. The offline data therefore require an event building step in order to assemble data frames from the various subsystems that correspond to a unique BNB spill. For ANNIE, the event building process relies primarily on matching timestamps. The central trigger stores a timestamp and identification code for every trigger it produces, including the undelayed copy of the beam trigger sent to the LAPPD readout electronics (LAPPD beam gate) and the delayed copies sent to the PMT (PMT trigger) and MRD readout electronics (MRD trigger).  
 
For the PMT system, a \pmtwin\/ readout window is recorded for every PMT in response to a beam trigger. If at least one tank PMT registers a hit of 7 photoelectrons or more in the initial \pmtwin\/ window, the readout window is extended to \pmtextwin\/ for all PMTs. In the first phase of the event building process, PMT readout windows (regardless of length) are matched to a CTC PMT trigger timestamp if the window start time is within 100 ns of the PMT trigger timestamp. This tolerance accounts for signal propagation delay and the response latency of the PMT electronics.

The MRD works similarly, although in this case the default readout window for each MRD paddle is \mrdwin\/. Readout windows for the MRD paddles are matched to the CTC MRD trigger timestamp if they are within 3 ms. The large tolerance here is due to the limited accuracy of the MRD readout window timestamp, which is taken from a computer system clock. 

The LAPPD system, as previously noted, timestamps the heartbeat, beam gate, and self-generated trigger signals using a 64-bit counter recording the 320 MHz ticks of the 40 MHz FPGA clock multiplied by eight. The sync frames recorded by each LAPPD ACDC can in principle be used to match the local LAPPD counter clock to the CTC global clock. However, there is an ambiguity as to which CTC heartbeat timestamp matches the first sync frame recorded by the LAPPD (the \emph{synchronization offset}). 
A distinct feature of the operations of the Fermilab accelerator complex provides the information needed to resolve this synchronization offset ambiguity. 

\begin{figure}[]
	\begin{center}
		\includegraphics[width=1.0 \linewidth]{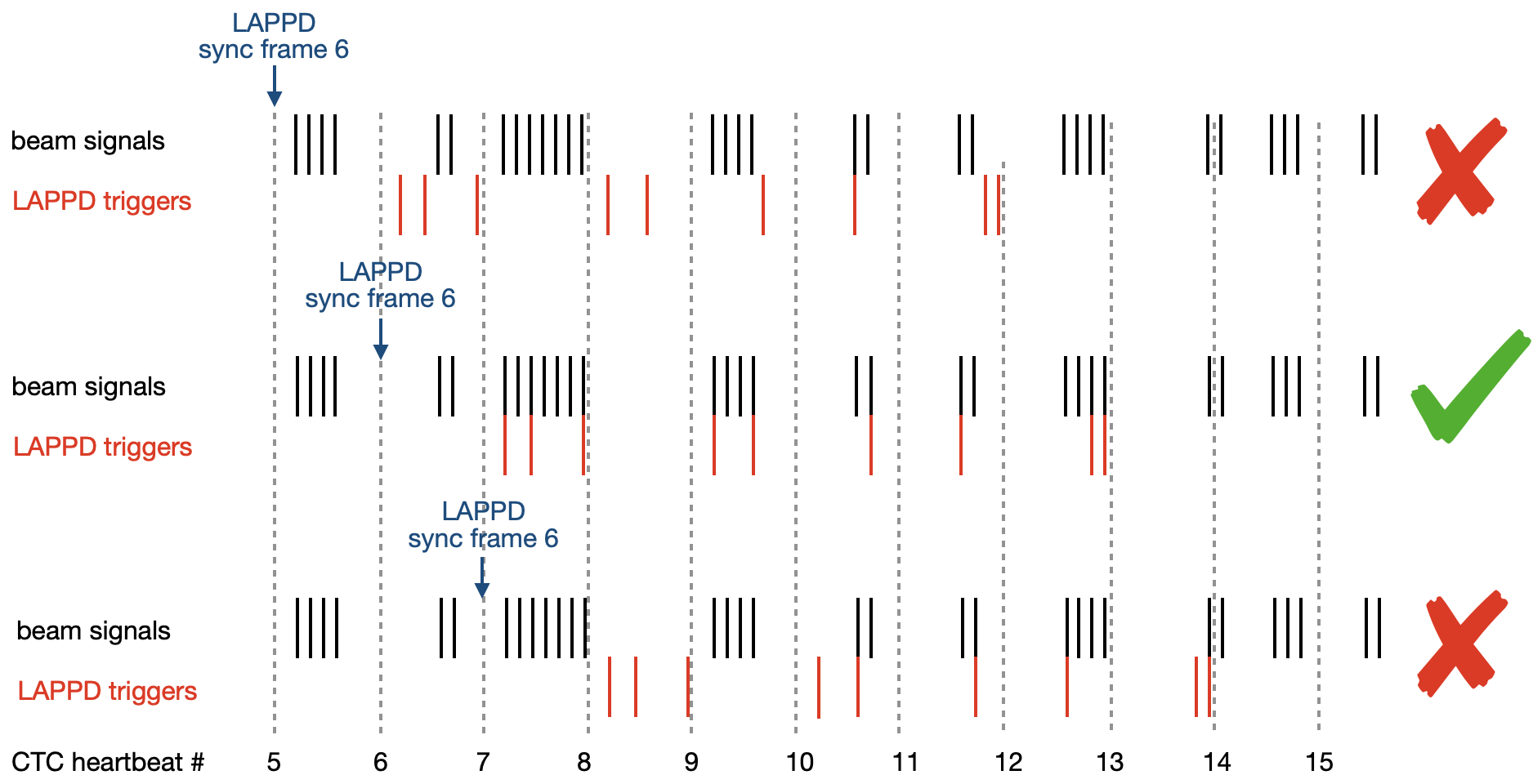}
	\end{center}
	\caption{A cartoon representation of the timing alignment used to correctly match LAPPD data frames to the rest of the corresponding beam events. The LAPPD receives a heartbeat signal from the CTC and records the local timestamp in a sync frame. There is an ambiguity as to which CTC heartbeat timestamp matches the first LAPPD sync frame. However, the time structure of the LAPPD neutrino triggers reflects the unique fingerprint of the beam time structure. Thus, when the correct LAPPD sync frame is aligned with the correct CTC sync frame, LAPPD triggers will be maximally and uniquely matched to beam signals in time. }\label{fig:beam_allignment}
\end{figure}

The Fermilab Booster provides protons not only to the BNB, but also to NuMI and the test beam facility (FTBF). The Booster complex's algorithm for allocating spills to the three beamlines means that the spills delivered to BNB alone irregularly sample the 15Hz RF on which the Booster operates. As a result, each recorded data file, (typically around \partfile\/ of data), is characterized by a distinct pattern of time separations between CTC beam gates. It is expected that the correct association between the LAPPD local counter and global CTC clock will correspond to the minimum average time separation between CTC beam gate timestamps and LAPPD local timestamps converted to global time. A fit iterates over all possible synchronization offsets for a given data segment to find the one that results in the minimum average time separation for all events. Each data segment is processed in conjunction with the preceding and subsequent segments in order to avoid data loss at file boundaries. Figure~\ref{fig:beam_allignment} shows a simple cartoon of this process.

A timing instability in the phase-locked loop (PLL) synchronization of the PAL with the central clock was identified during 2023 and fully resolved with a hardware update in September of that year. All data collected since that update exhibit stable timing and require no additional treatment. The data presented in this paper, however, were acquired prior to the fix and therefore require an additional software-based correction to compensate for clock drift. 
After applying the synchronization offset fit and the secondary correction, any LAPPD data frame occurring within $400~\mu$s of the central beam time can be reliably matched to the rest of the system. This is loose enough to account for both the programmed signal delay applied by the ACC and residual uncertainties in the LAPPD-CTC timing synchronization. 

The PMT trigger, MRD trigger, and LAPPD beam gate are all issued when the CTC recognizes that a beam spill is imminent. 
The CTC timestamps each of these three signals and records them in its data stream. It is straightforward to associate these three timestamps with one another. Each subsystem also timestamps the relevant signal upon receipt with the subsystem's local clock and includes this information in the subsystem data stream. 

By matching the individual subsystem timestamp with the corresponding CTC timestamp, data frames from the individual subsystems can be assembled into a singular record of a beam spill in the ANNIE detector, also referred to as a beam event.

\begin{figure}[]
	\begin{center}
		\includegraphics[width=1.0 \linewidth]{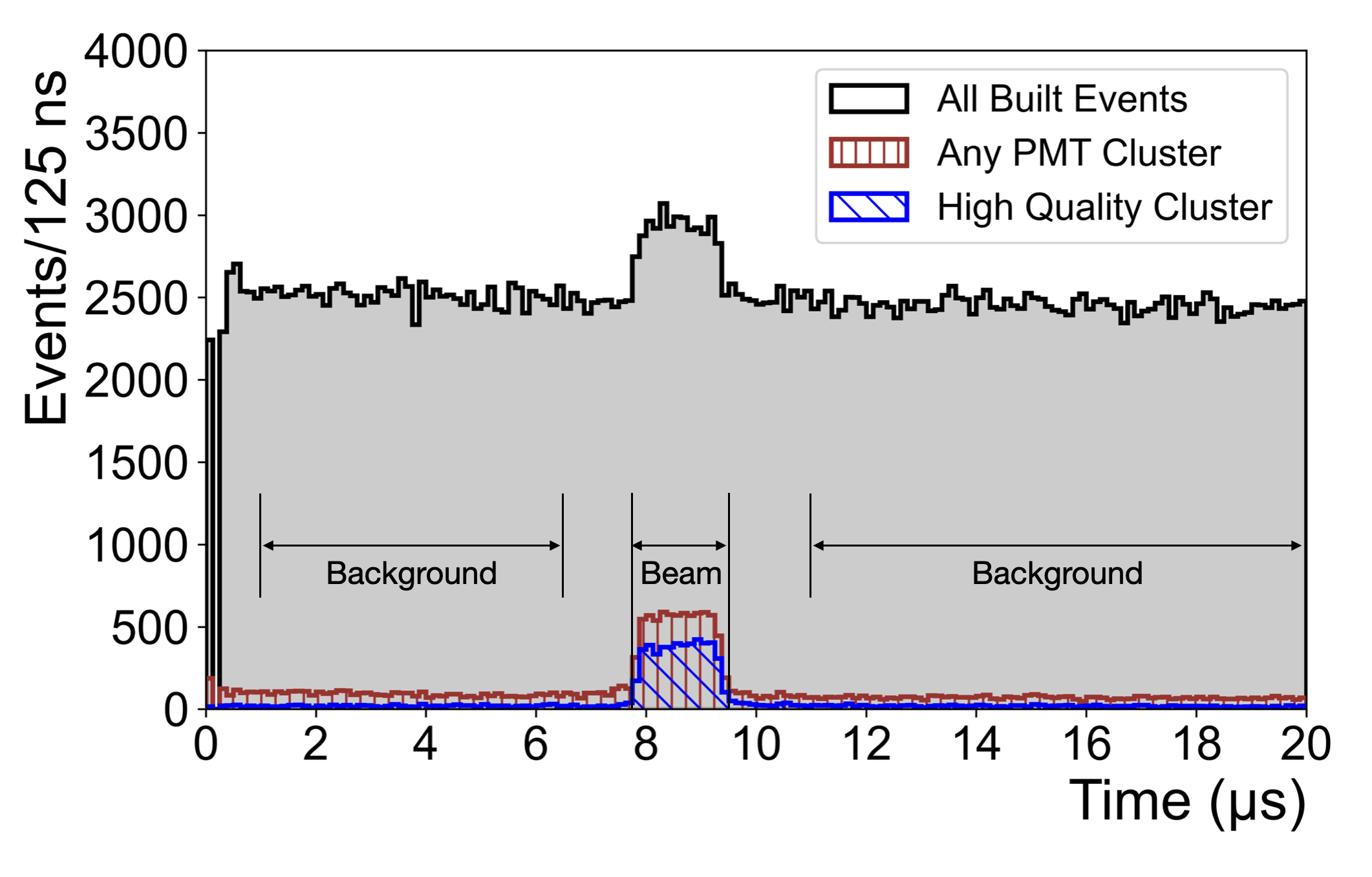}
	\end{center}
	\caption{Time difference, in microseconds, between the receipt of the beam gate by the \pal\/ and the LAPPD autonomous trigger, for a sample of over 400k ANNIE beam events (black-gray) where data from an LAPPD has been matched to a PMT readout window.  The central peak corresponds in width to the length of the beam spill.  The bracketed regions indicate the regions used to estimate off-spill background and select beam events. The subset of events that pass 
    a \emph{PMT cluster} requirement of five or more PMT hits within 50 ns is overlaid in red (striped fill). The subset of these events with high-quality PMT clusters is shown in  blue (cross-hatch).
 }
		\label{fig:raw_beam_timing}
\end{figure}

Figure \ref{fig:raw_beam_timing} shows, for the full sample of beam events, the difference between the time at which the LAPPD received the beam gate and the time at which the LAPPD self-triggered. The structure of this distribution shows a central peak around 1.6 $\mu$s wide, corresponding to the beam spill, with a substantial (${\sim}1150$ events$/125$ ns) and relatively flat background to either side. That background is a combination of dark and electronics noise triggers, cosmic ray muons, and beam-related backgrounds arising during the spill. The presence of a peak consistent with the beam spill indicates that the LAPPD is in fact detecting events correlated with the beam. 

\subsection{Selection of Charged-Current Neutrino Interactions}

In this section, we consider the impact of cuts using information from the entire ANNIE detector that are designed to select charged-current neutrino interactions either in the tank or in the dirt and rock upstream of the detector (``dirt muons''). 
In evaluating the impact of these cuts on the timing distribution of the matched LAPPD data frames, we see that these cuts preferentially select data frames in time with the spill. An important quality metric for each cut is the sample purity within the beam spill region ($7.75 -9.35$ $\mu$s), that is, the number of events with an LAPPD data frame in the spill region after subtracting the estimated background, divided by the total number of events with an LAPPD data frame in the spill region. We estimate the background rate using two off-beam regions, 1-6 $\mu$s and 11-20 $\mu$s. Both the selected spill region and the background estimation regions are shown in Fig. \ref{fig:raw_beam_timing}.

The red (striped fill) overlay in Figure~\ref{fig:raw_beam_timing} shows the sub-sample of  matched beam events that contain at least one \emph{PMT cluster} within the initial \pmtwin\/ readout window. A PMT cluster represents a set of PMTs that registered light consistent with a single track, shower, or neutron capture gamma cascade. It is defined as a set of at least five PMT hits within 50 ns of one another (50 ns being approximately twice the maximal expected time difference between hits due to light propagation times within the tank). Clusters are found by searching the full $70~{\mu}s$ readout for the 50 ns window that contains the maximal number of hits. These hits are assigned to a cluster and removed from consideration. The process is then repeated until a full scan of the $70~{\mu}s$ readout fails to produce five or more hits in any 50 ns window. Selecting only PMT clusters for which the average hit time lies in the initial readout window isolates beam-induced neutrino interactions and eliminates PMT clusters due to later detector activity, such as neutron interactions and muon decays. Here, the term hit is defined based on the techniques for baseline subtraction, pulse finding, and waveform calibration described in Ref~\cite{Back_2020} for tank PMTs.

This PMT cluster requirement eliminates the majority of the uncorrelated background while sparing the beam events. In fact, as shown in Table \ref{tab:CutTable}, the estimated efficiency of this cut for true beam events is statistically consistent with $100\%$. The effectiveness of this cut at suppressing background for the LAPPD data frames indicates that the incorporation of LAPPD data into ANNIE events is reliable and robust. 

The PMT cluster requirement is even more effective if we consider only ``high-quality'' PMT clusters where the cluster charge is distributed relatively evenly between hits, rather than coming primarily from one or two PMTs. This is implemented by requiring that 
\begin{align}
    \sqrt{\frac{\sum_{i=1}^N {Q_i}^2}{\left(\sum_{i=1}^N {Q_i}\right)^2} - \frac{1}{N}} < 0.2
\end{align}
where $Q_i$ is the integrated charge seen by the ith PMT and $N$ is the number of operational PMTs. The $Q_i$ for a PMT not in the cluster is treated as zero. This requirement is adapted from the WATCHBOY experiment \cite{DAZELEY2016151} and suppresses clusters arising from instrumentation effects and noise. The additional impact of the PMT cluster quality cut is illustrated in the blue (cross-hatch) overlay in Figure~\ref{fig:raw_beam_timing}.

\begin{figure}[]
	\begin{center}
		\includegraphics[width=1.0 \linewidth]{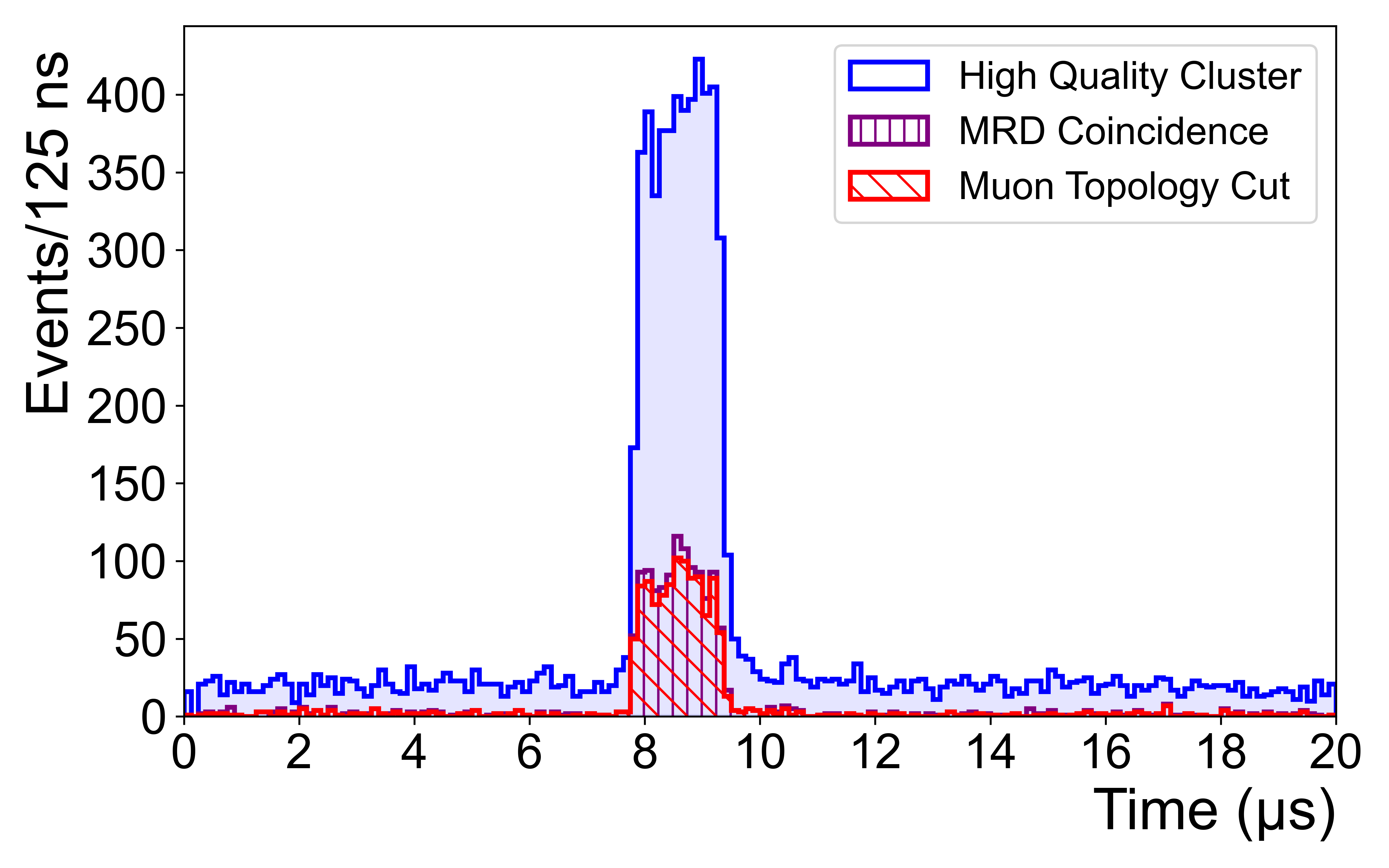}
	\end{center}
	\caption{Time difference, in microseconds, between the receipt of the beam gate by the \pal\/ and the LAPPD autonomous trigger,
    for all LAPPD data frames paired with a high-quality PMT cluster (blue, solid fill), all LAPPD data frames matched to both a PMT cluster and an MRD track (violet, vertical stripe), and all LAPPD data frames matched to a PMT cluster and MRD track that have no activity in the forward veto (red, cross-hatch).}\label{fig:beam_timing_cuts}
\end{figure}

Figure~\ref{fig:beam_timing_cuts} shows the further impact of requiring activity in the MRD consistent with a muon arising from a charged-current neutrino interaction. TDC units attached to the MRD scintillator paddles register the time at which a muon has passed through the paddle and produced scintillation light, i.e. an MRD hit. A cluster of at least four MRD hits within 30 ns, two from vertical paddles and two from horizontal paddles, is required to successfully reconstruct a muon track. The muon track reconstruction procedure used for the MRD is discussed in greater detail in Refs~\cite{Fischer:2020hjj} and \cite{NieslonyThesis}. This cut has the most dramatic impact on the available statistics, since it limits the sample to forward-directed muons energetic enough to penetrate the MRD. In order to select a sample of muons that we also expect to illuminate the LAPPD, we further require that the Cherenkov disk observed by the tank PMTs overlaps with or surrounds the LAPPD (muon topology cut).

\begin{table}[ht]
\footnotesize
\centering
\caption{The total number of LAPPD data frames in (left to right) the $20 {\mu}s$ window after the beam gate, the total number of LAPPD data frames restricted to the $1.6 {\mu}s$ window of the beam spill, and the estimated number of true beam events after each of the following sequence of cuts: LAPPD to PMT time tolerance < 100$\mu s$, at least 5 PMT hits in a 50ns window, MRD activity, requiring that the Cherenkov disk in the tank overlap the LAPPD, and requiring that there be no hits in the FMV. The number of true beam events is obtained by subtracting the background (estimated using the average event count outside of the 6 to 12 $\mu s$ window) from the total number of events within the $1.6 \mu$s beam peak. Sample purity is obtained by taking the ratio of the estimated number of beam events to the number of events in the $1.6 {\mu}s$ spill window.}
\begin{tabular}{ccccc}
\toprule
\textbf{Cut} & \textbf{Events [20 ${\mu}$s]} & \textbf{Events [1.6 ${\mu}s$]} & \textbf{Beam Events (est.)}  &\textbf{Purity (\%)}\\
\midrule
Paired & 400024 & 38031 & 6126 $\pm$ 75.9 &  16.11 $\pm$ 0.200 \\
Any PMT cluster & 19265 & 7047 & 6126 $\pm$ 12.9 &   86.94 $\pm$ 0.180  \\

High-quality PMT cluster  & 7902 \quad & 4737\quad & 4480 $\pm$ 6.8\quad & 94.58 $\pm$ 0.14\quad   \\

MRD track reconstructed & 1556 & 1190 & 1165 $\pm$ 2.1 & 97.86 $\pm$ 0.18  \\

In-time MRD coincidence & 1458 & 1133 & 1110 $\pm$ 2.0 &  97.96 $\pm$ 0.18\\

Muon topology cut, inclusive &1273 \quad & 1045\quad & 1030 $\pm$ 1.7\quad & 98.5 $\pm$ 0.2\quad   \\
No hit in FMV  &545 \quad & 457\quad & 451 $\pm$ 1.1\quad & 98.7 $\pm$ 0.2\quad   \\
\bottomrule
\end{tabular}
\label{tab:CutTable}
\end{table}

Table~\ref{tab:CutTable} reports the event yield after each selection cut for the entire sample and for the 1.6 ${\mu}$s spill region, as well as the estimated event yield for true beam interactions. From this table, we see that the additional background suppression provided by the MRD requirement is a modest (${\sim}3$ percentage points in signal purity) improvement over the high-quality PMT cluster requirement alone. Further requiring that the PMT and MRD clusters are within $750 \pm 50~\mathrm{ns} $ of one another provides another slight boost in signal purity by suppressing random coincidences between beam events and cosmic ray muons.
The muon topology cut improves signal purity by an additional percentage point. 

These cuts clearly select a high purity sample of muons arising from charged-current neutrino interactions in time with the BNB neutrino beam that illuminate the central ANNIE LAPPD. This not only establishes the first observation of beam neutrinos with an LAPPD, but also shows that we can leverage the full detector capabilities of ANNIE to select a useful calibration sample. With this sample, the response of the LAPPD can be studied in relation to the parameters of a muon neutrino interaction as reconstructed independently by other detector systems. 
We can select a subset of events useful for physics measurements by requiring the neutrino interactions take place inside the ANNIE detector volume. We do so by requiring that there be no activity in the FMV within the time window $750$ ns $< \Delta{t}_{FMV, PMT} < 850$ ns. As expected, this cut has no significant impact on signal purity. 

\subsection{A First Look at the Single-Muon LAPPD Response}

\begin{figure}[tp]
	\begin{center}
		\includegraphics[width=0.9\textwidth]{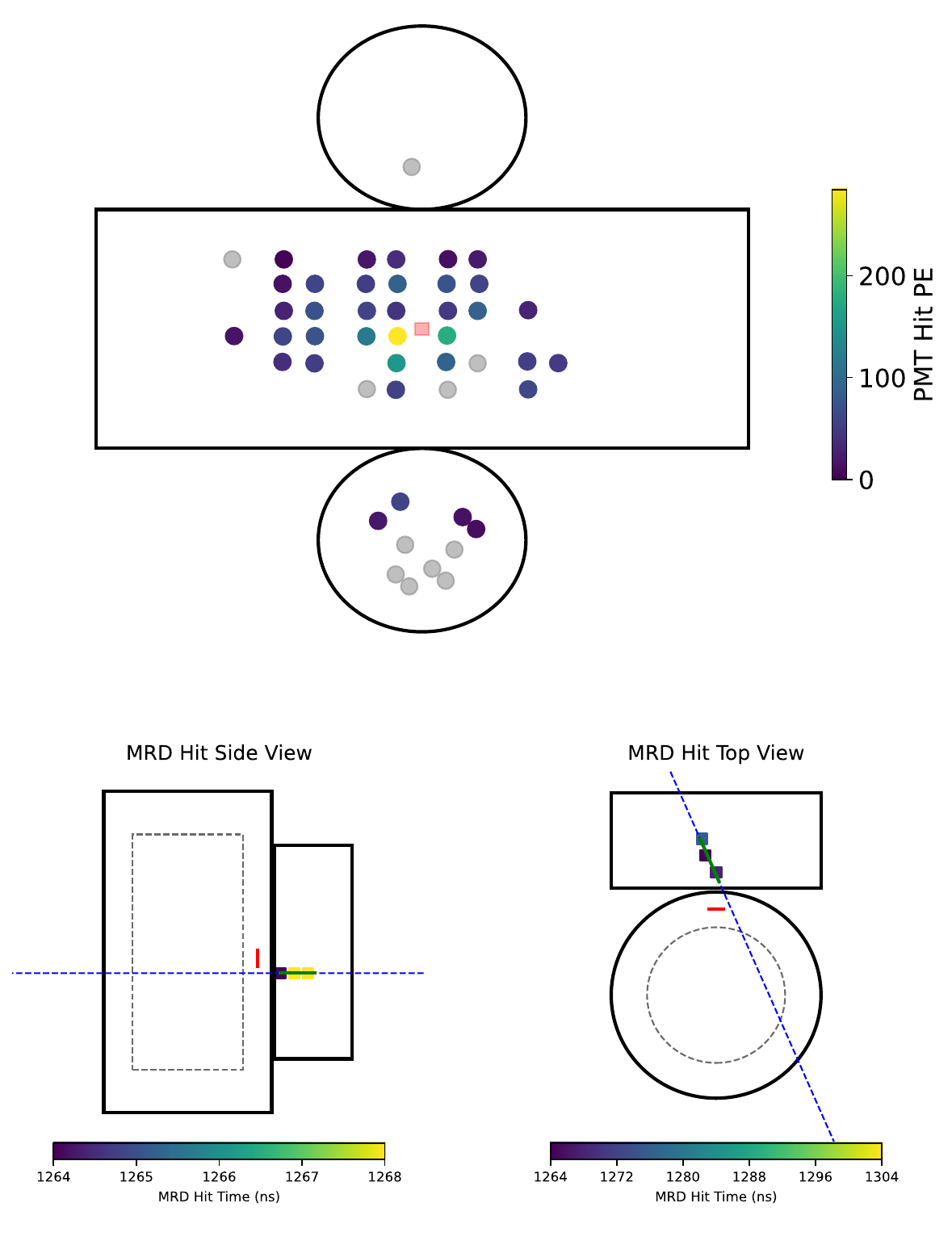}
	\end{center}	\caption{An event display showing the Cherenkov light pattern of a typical through-going muon event passing all selection cuts and captured by the central LAPPD, with the cylindrical detector unfolded and centered on the neutrino beam. The corresponding hits in the top and side views of the MRD are shown as colored blocks, while the projected muon track is shown as a dashed blue line. PMT hits above threshold are shown with colored dots, and non-operational PMTs are shown in gray. PMTs with low light levels are suppressed. The red square shows the position of the central LAPPD.}\label{fig:event56_eventdisplay}
\end{figure}

\begin{figure}
\begin{floatrow}
\centering
\subfloat[]{
\includegraphics[width=0.46\textwidth,trim={0cm, 0cm, 0cm, 0cm},clip]{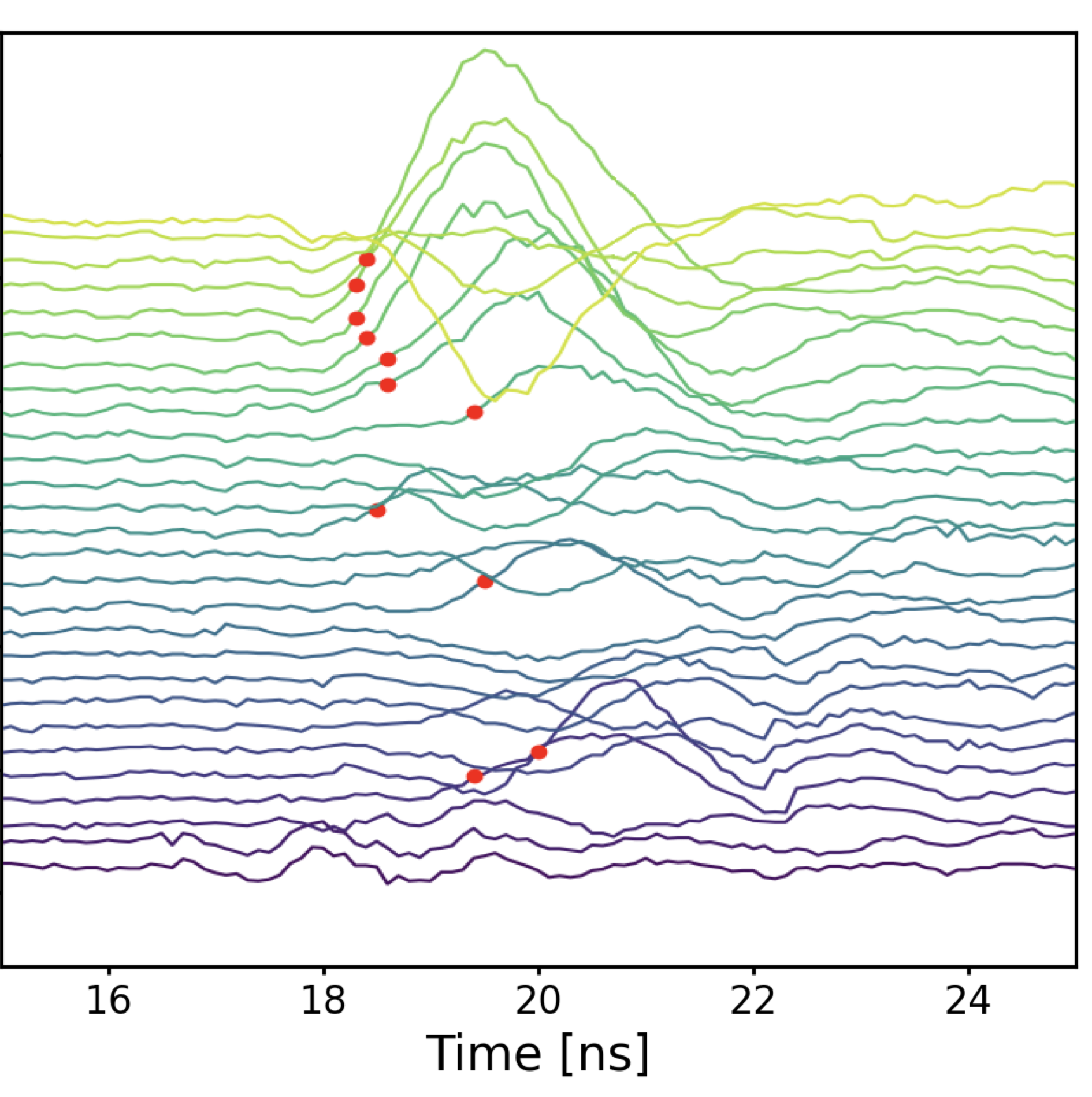}
}
\subfloat[]{
\includegraphics[width=0.46\textwidth,trim={0cm, 0cm, 0cm, 0cm},clip]{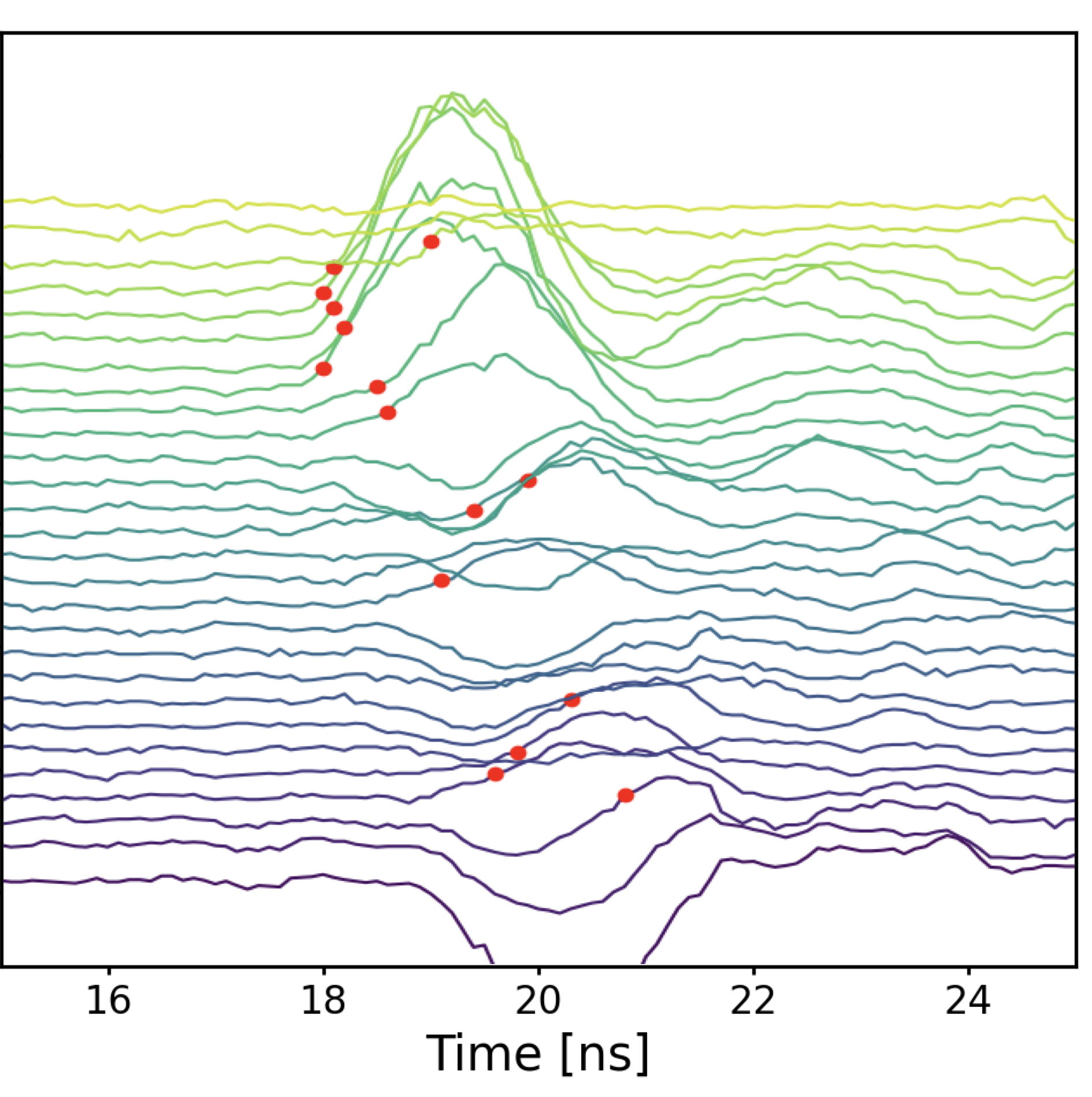}
}
\end{floatrow}
\qquad
\subfloat[]{
        \centering        \includegraphics[width=1.0\textwidth]{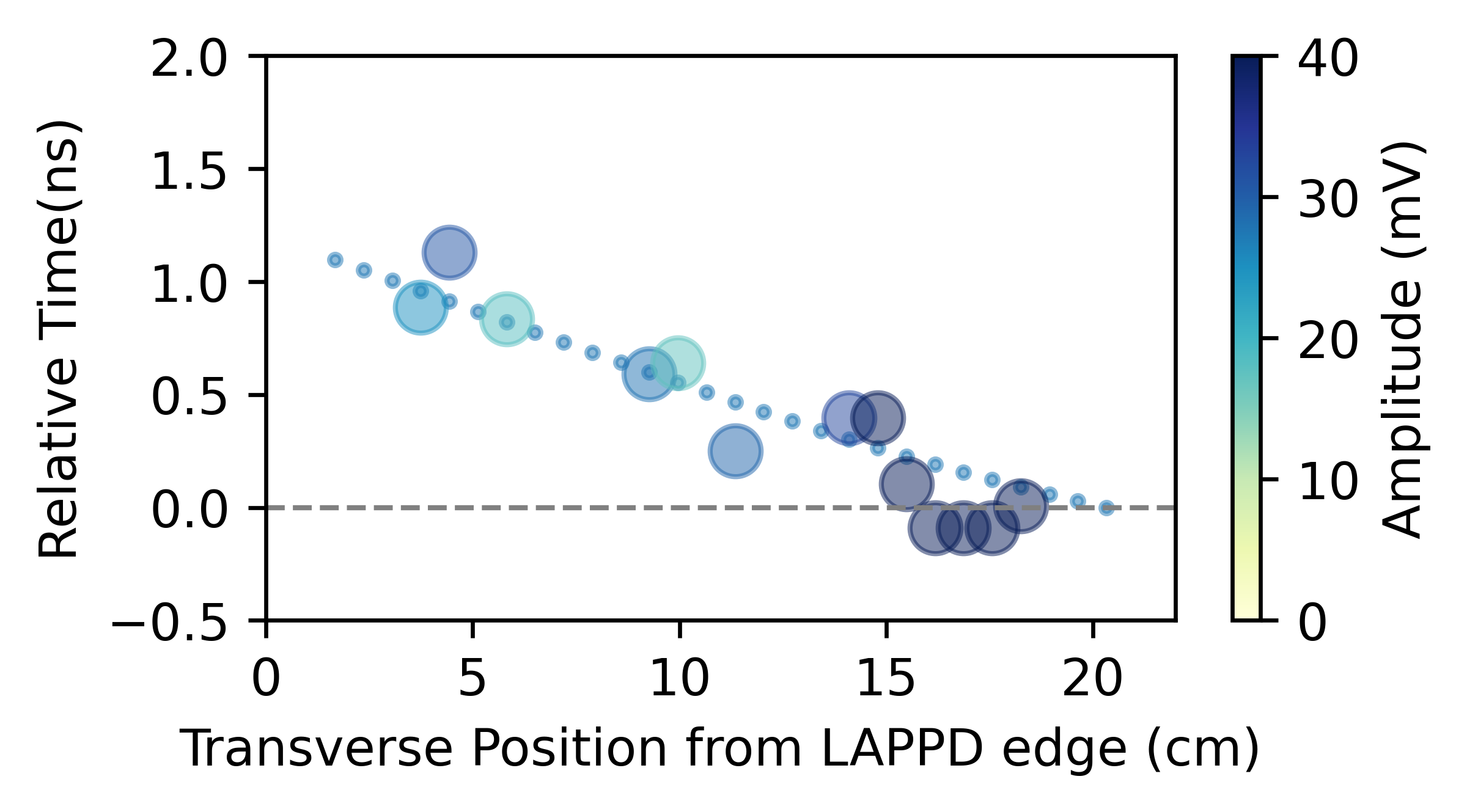}
}
	\caption{Top: Waveforms from both sides of the LAPPD microstrip readout, corresponding to the event shown in Figure \ref{fig:event56_eventdisplay}. The estimated arrival time for each waveform is shown as a red dot. Bottom: Comparison of the average arrival time of the multi-photoelectron-induced pulses on each LAPPD strip (large colored circles, with the color scale indicating hit amplitude in mV) to the predicted arrival times inferred from the MRD track (small blue dots). } 
	\label{fig:event56_muonlappd}
\end{figure}

Figures~\ref{fig:event56_eventdisplay} and \ref{fig:event56_muonlappd} show the ANNIE event display and LAPPD response for a dirt muon event that passes all other selection cuts, with the reconstructed muon track passing close to the central LAPPD. Figure~\ref{fig:event56_eventdisplay} shows
that the MRD information is consistent with a horizontal muon exiting the tank slightly below and to the left of the LAPPD. The distribution of charge in the PMT cluster is consistent with a Cherenkov disk overlapping the LAPPD. The top panel of Figure \ref{fig:event56_muonlappd} displays the raw LAPPD waveforms from this event. As previously noted, signals from both sides of each anode microstrip are read out and digitized, resulting in a total of 56 digitized waveforms for 28 microstrips.

As noted in Section~\ref{sec:intro}, LAPPDs are {\it imaging} photosensors, i.e. they register information about the spatio-temporal pattern of light arriving at the LAPPD surface.
In order to give a heuristic impression of the type of entangled time and position information provided by the LAPPD waveforms, we estimated the arrival time of every pulse as the time when that pulse exceeded a 10 mV threshold (shown as red dots in Figure \ref{fig:event56_muonlappd}).  For each microstrip, we averaged the pulse arrival time between the two sides and plotted it as a function of distance from the right edge of the LAPPD in the plane of the detector window. As the strips are oriented vertically in the ANNIE tank, the microstrips can be used to estimate distance from the LAPPD edge in the plane transverse to the beam. 

Figure \ref{fig:event56_muonlappd} shows a ${\sim}1$ ns time gradient in the arrival-time of the light across the surface of the LAPPD. As expected, early-arriving light produces higher-amplitude pulses. Using ray-tracing of Cherenkov photons, we can also predict the arrival time gradient of light at the surface of the LAPPD for a muon with the track parameters reconstructed using only information from the ANNIE MRD. The reconstructed gradient appears consistent with the prediction. This suggests that an LAPPD collects enough information to at least partially constrain the track parameters of a muon---something that is not possible with a single conventional photosensor.

We have observed that most visible waveforms in ANNIE LAPPD events do not correspond to a single photoelectron liberated by the photocathode, but are rather a superposition of many such pulses. This superposition arises as Cherenkov photons arrive within a time period comparable to the characteristic width of a single-photoelectron pulse in time; moreover, the charge clouds induced by these photoelectrons can spread over multiple microstrips. These features of the data suggest that reconstruction is better done by fitting a holistic model of the voltage-time response rather than through deconvolution and isolation of individual photons. Future work will develop both the fitting method and the necessary modeling of the time-voltage response for different muon parameters in order to benchmark these capabilities.

\section{Conclusions}
\label{sec:conclusion}

ANNIE has achieved the first successful deployment and sustained operation of a Large Area Picosecond Photodetector (LAPPD) in a running neutrino experiment, and has recorded the first light from beam neutrino interactions with an LAPPD. These results move LAPPDs from test-stand and simulation studies into demonstrated experimental use.

By combining LAPPD data with information from all other ANNIE sub-detector systems, we selected a high-purity sample (98\% pure) of beam neutrino interactions (1,045 events in total) that illuminated the LAPPD. This sample, which includes accompanying MRD tracks, provides a realistic testbed for developing and validating LAPPD-based track reconstruction. The waveform morphologies and timing gradients shown here demonstrate that a single LAPPD can deliver rich spatio-temporal information with power to constrain muon track parameters.

The early deployments have also yielded essential system-engineering insights. ANNIE provides the first experience with long-term operation of LAPPDs in a space-constrained water detector and has already informed improvements to slow controls, waterproof housings, and front-end readout electronics. These lessons are directly transferable to future multi-module systems.

With additional LAPPDs and an extended run, ANNIE has collected an order of magnitude more neutrino interactions with greater geometric coverage. That dataset will enable quantitative reconstruction studies and performance benchmarks that are not yet possible with the dataset presented here.

More broadly, ANNIE demonstrates that LAPPDs can change how we approach neutrino detection with water-based targets.
The combination of picosecond timing and millimeter-scale spatial sampling opens a path to enhanced physics reach, new reconstruction techniques, and novel detector concepts for next-generation neutrino and collider experiments.

\begin{appendices}

\section{Surface-to-Water Cabling}
\label{app:AppC}

As discussed in Section~\ref{sec:Technical}, 
high-density waterproof cables from Falmat (Falmat XtremeNet) \cite{Falmat} connect the surface rack electronics and power to the \pal. Here we describe in detail the design of this cable system.

Each \pal\/ takes two waterproof cables, and each cable carries 4 twisted-pair 23 AWG data lines and four singular 18 AWG control lines. One of the four data lines carries copies of the central clock and heartbeat signals, one carries a \emph{beam gate} signal provided by the ANNIE central trigger, and the remaining two support the data streams that carry readout configuration commands from the surface to the water and return data from the water to the surface. The four control lines are used in pairs for power and ground, a thermistor, a catastrophic leak sensor, and CANBUS communications as described in Appendix~\ref{app:AppA}. 

The surface end of the waterproof cable terminates the Cat6 twisted-pair data lines in a female RJ-45 jack. A Cat6 cable is used to patch the data lines directly from this jack to the surface components of the LAPPD readout electronics. This maintains the native noise and cross-talk performance of the cables. The power and control lines go to an Amphenol Industrial PT06E-14-12P connector \cite{Amphenol} that mates to a connector on the ``breakout boxes'' that serve power and slow controls at the surface.

The cables are routed to the \pal\/ through rectangular openings (LAPPD deployment ports) at the top of the tank. The \pal\/ end of the Falmat cable is spliced to a waterproof lead cable attached to one half of a floodproof SubConn Power Ethernet connector from MacArtney~\cite{SubConn}. The mating half of the SubConn connector is installed in the LAPPD housing. The Falmat XtremeNet cable itself is compatible with Gd-water and is left exposed, but the SubConn connector and lead cable are not as Gd-compatible and have been Teflon-wrapped to prevent leaching.

Inside the \pal\/, the data line twisted pairs exit the SubConn connector via an RJ-45 Cat5e patch cable. The power and ground lines are directly routed to terminal blocks on the LVHV.

\section{Power Distribution and Slow Controls}
\label{app:AppA}

As noted in Section~\ref{sec:Technical}, the LAPPD subsystem has its own dedicated power supply and distribution system. An Excelsys Xgen XLC\cite{Excelsys} power supply system provides the primary power source for the LAPPD subsystem, with two 12V 240W Xg3 modules powering the {\pal}s high- and low-voltage distribution systems. A single 5V 200W Xg2 module powers the ACCs and Raspberry Pis on the surface. 

\begin{figure}[]
	\begin{center}
		\includegraphics[width=1.0 \linewidth]{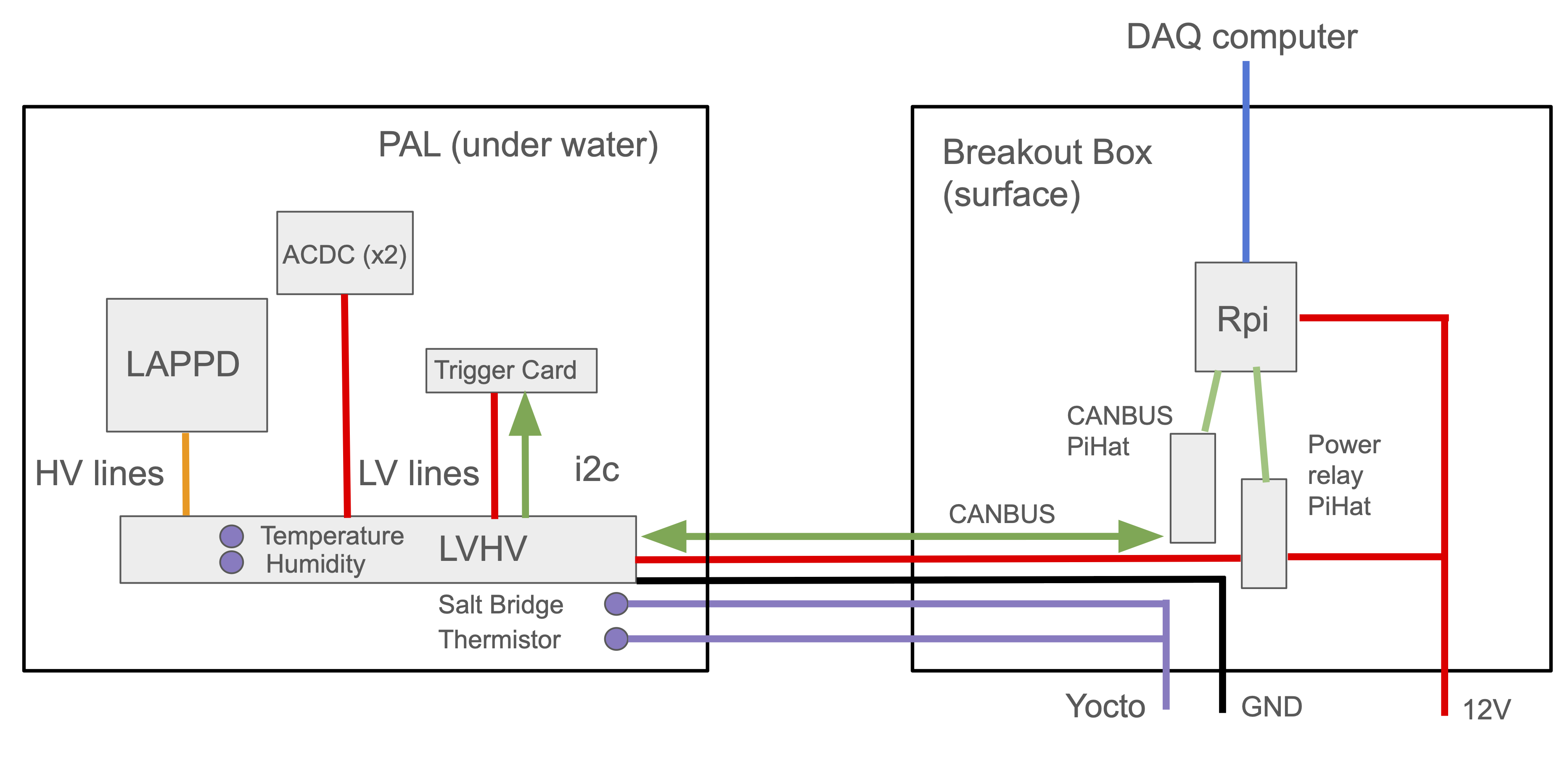}
	\end{center}
	\caption{Schematic of the power distribution and slow controls on the surface and underwater. A single ground and 12V power line are connected through one of the two water proof cables where they connect to the LVHV inside the {\pal}. The LVHV provides three low voltages to the ACDC cards (3.3V, 3.2V, 1.8V) and one low-voltage line to the trigger card (2.5V). The LVHV also provides the 5 high voltages to the LAPPD. The LVHV sets trigger card values through an i2c interface, and it communicates with the surface controls via CANBUS. The salt-bridge used to detect catastrophic leaks is directly read out on the surface. Other humidity and temperature sensors are read out by the LVHV.}
		\label{fig:slowpower}
\end{figure}

Power is routed and delivered through breakout boxes, which (a) serve power to the underwater {\pal\/} modules, (b) receive  alarm and monitoring signals from the {\pal}s, and (c) contain two model B$+$ Raspberry Pi 3s~\cite{RPi} that manage slow controls, trigger configuration, and alarms. A single breakout box is designed to service two {\pal\/}s. In addition to providing an interface for the termination of the waterproof cables and distributing power and communications lines, the Raspberry Pis serve as the interface between the DAQ and LAPPD slow controls. Each Raspberry Pi is outfitted with two add-on mezzanine cards (Pi HATs): a Copperhill technologies PiCAN2 CANBUS interface\cite{Copperhill} for communications with the underwater slow controls on the LVHV board and an Electronics-Salon RPi Power Relay Board Expansion Module~\cite{Salon} for shutting off the power to the {\pal\/} in the event of a serious water leak. 
A unit cell of the breakout box and how it maps to the power and slow controls of the {\pal\/} is shown in Figure~\ref{fig:slowpower}.

The internal readout and trigger system require a variety of low ($\leq 12$V) voltages. Power for the \pal\/ is brought in on a single pair of lines (12V and ground) from the surface. All other needed voltages are produced inside the module, on a board called the {LVHV board}. In addition to voltage regulation and distribution, the LVHV board also interfaces with environmental sensors in the module and serves as an AC-coupled pass-through for the twisted pair communications with the surface electronics.

The ACDC readout board for each LAPPD require three independent low voltages: 3.4V, 3.2V, and 1.8V. Rather than produce these voltages onboard, which would generate heat too close to the LAPPD, these 3 voltages are made on the LVHV board and delivered each ACDC on three separate lines. Linear regulators on the ACDC cards help to stabilize any transient behavior on the low voltage lines.

Control and monitoring of the LVHV board is provided by a Microchip Inc dsPIC33EV256GM106 (PIC) microcontroller\cite{PIC}. The PIC can control and monitor the low voltage and high voltage circuits on the LVHV, read out onboard humidity and temperature sensors, set thresholds on the trigger board, and communicate with a Raspberry Pi on the surface electronics using a CANBUS interface.

\begin{figure}[]
	\begin{center}
		\includegraphics[width=0.5 \linewidth]{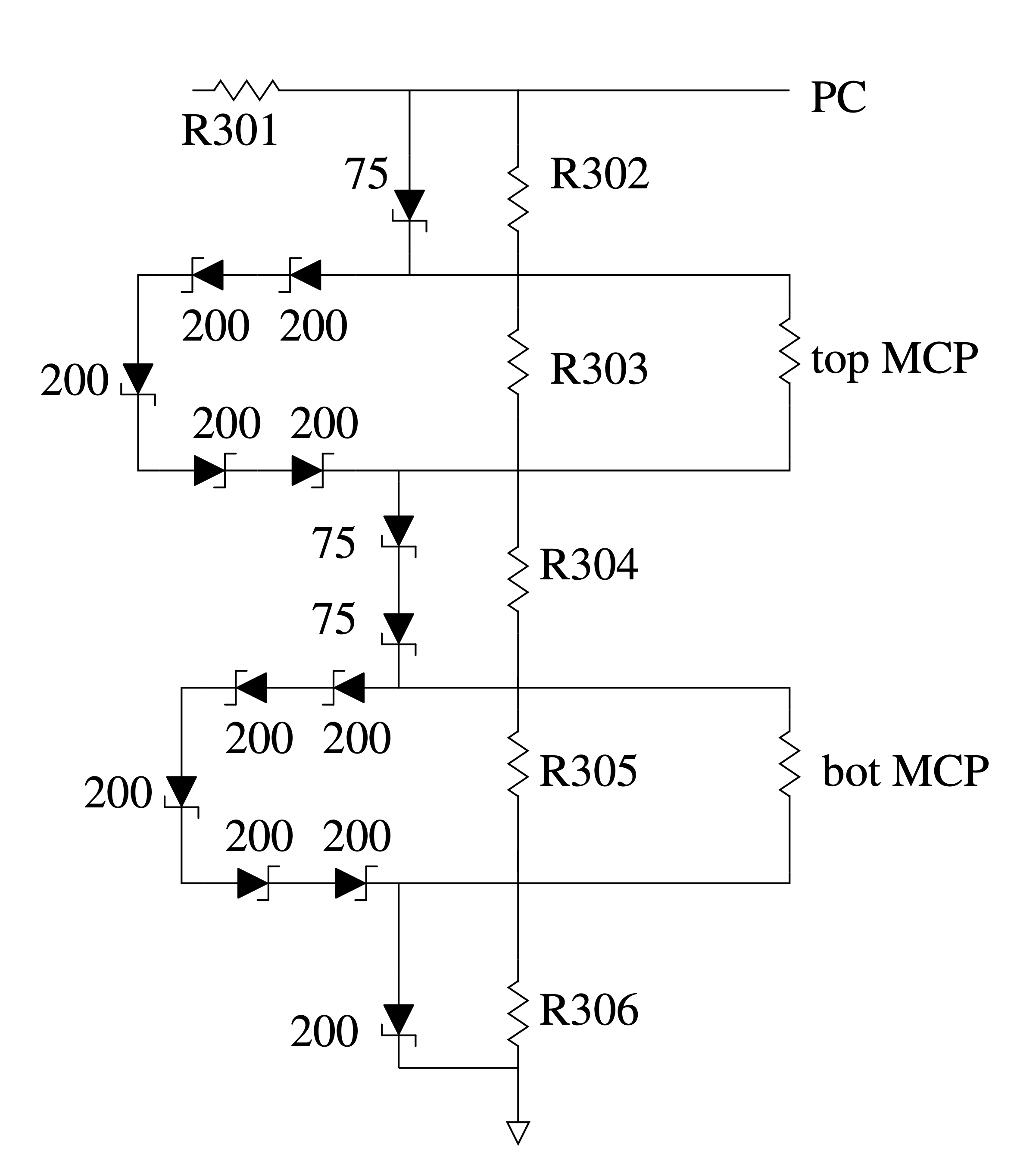}
	\end{center}
	\caption{Schematic of the high-voltage divider with zener diodes used to power the five stages of the LAPPD (photocathode, top of entry MCP, bottom of entry MCP, top of exit MCP, bottom of exit MCP).}
		\label{fig:hv_schematic}
\end{figure}

Operation of an ANNIE LAPPD requires a chain of 5 high voltages spanning 3kV-0V.
The maximum high voltage for the LAPPD is generated by a single EMCO C40N  power converter\cite{EMCOC40N}. This voltage is divided down by a resistive divider network, with parallel zener diodes used to protect against overvoltages across any single stage of the LAPPD. The individual MCPs of the LAPPD are connected with the resistive divider in series. If the individual plate resistances are mismatched, a parallel surface mount resistor is used to equalize the equivalent resistance. Figure~\ref{fig:hv_schematic} shows a schematic of the setup. A single sensor on the LVHV allows read back of a test point that reflects the high voltage value at the top of the stack divided by ten. However, we are unable to interrogate the voltages of each stage independently on the deployed LAPPD {\it in situ}.

\textbf{}

\end{appendices}

\acknowledgments

A very special thank you to Paul Rubinov at the Fermilab Electronics Design group for providing essential resources and personnel for the LAPPD effort, and personally advising the collaboration on a number of important electronics issues. We could not have prepared the LAPPDs or made the significant and iterative changes to the LAPPD electronics necessary for deployment without keen eyes and careful hands of Albert Dyer (FNAL). Paul Stucky at UC Davis was essential to completing the design and execution the LVHV boards. David Greenshields (Warwick University) was critical in developing, testing, and maintaining the LAPPD firmware. Henry Frisch, Evan Angelico, Eric Oberla, Mircea Bogdan, and Mary Heintz at U Chicago provided valuable experience and expertise on the UC-designed PSEC electronics. Thank you to the entire team at Incom Inc, not only in making LAPPDs a reality, but also for their close collaboration and quick feedback in the effort to see them deployed.

The ANNIE collaboration would like to honor the memory of Gary Varner at the University of Hawaii, who served as a guide, mentor, and friend to so many of our collaborators, and whose vast expertise was critical in the work preceding ANNIE and in guiding our thinking about our timing synchronization.

This work was supported by: (1) the U.S. Department of Energy (DOE) Office of Science under award number DE-SC0009999 (UC Davis), award number DE-SC0015684 (Iowa State, award number DE-SC0024684 (Florida State University), and DE-SC0014223 (South Dakota School of Mines),  and by the DOE National Nuclear Security Administration through the Nuclear Science and Security Consortium under award number DE-NA0003180 (UC Davis), (2) Deutsche Forschungsgemeinschaft grants 456139317 (Hamburg) and 490717455 (Mainz and T\"ubingen), and 552099472 and RTG2796 ``Particle Detectors'' (Mainz) (3) National Science Foundation Grant No. PHY-2310018 (Ohio State), PHY-2047665 (Rutgers), and OIA-2132223 (South Dakota Mines), (4) UK Research and Innovation FLF MR/S032843/1 (Warwick), (5) Scientific Research Projects (BAP) of Erciyes University under the grant numbers of FBAU-2023-12325, FDS-2021-11525 and FBA-2022-12207.
Work conducted at Lawrence Berkeley National Laboratory was performed under the auspices of the U.S. Department of Energy under Contract DE-AC02-05CH11231.
Work conducted at Lawrence Livermore National Laboratory was performed under the auspices of the U.S. Department of Energy by Lawrence Livermore National Laboratory under contract DE-AC52-07NA27344, release number LLNL-JRNL-2009446. The work conducted at Brookhaven National Laboratory was supported by the U.S. Department of Energy under contract DE-AC02-98CH10886. The project was supported by the U.S. Department of Energy, National Nuclear Security Administration, Office of Defense Nuclear Nonproliferation Research and Development (DNN R\&D). Finally, we gratefully acknowledge all the Fermilab scientists and staff who supported this work through their technical expertise and operational assistance at the Booster Neutrino Beam.

\clearpage

\bibliographystyle{model1-num-names}
\bibliography{LAPPD}

\end{document}